\pdfoutput=1 
\documentclass[a4paper,UKenglish,cleveref, autoref, thm-restate]{lipics-v2021}

\usepackage[utf8]{inputenc}
\usepackage{float}
\usepackage{listings}
\usepackage{xcolor}
\usepackage{booktabs}
\usepackage{tabularx}
\usepackage{array}
\usepackage{caption}
\usepackage{subcaption}
\usepackage{tikz}
\usepackage{wrapfig}
\usepackage{enumitem}
\usetikzlibrary{positioning, shapes.geometric, arrows.meta, calc, fit, backgrounds}

\definecolor{riscvblue}{RGB}{0, 51, 153}
\definecolor{riscvgray}{RGB}{100, 100, 100}
\definecolor{coqpurple}{RGB}{127, 0, 85}
\definecolor{coqgreen}{RGB}{0, 128, 0}

\lstdefinelanguage{RISCV}{
    keywords={lui, sub, auipc, addi, jalr, slli, add, lw},
    sensitive=false,
    comment=[l]{\#},
    morecomment=[s]{/*}{*/},
    keywordstyle=\color{riscvblue}\bfseries,
    commentstyle=\color{riscvgray}\ttfamily
}

\definecolor{primary}       {RGB}{   0,  48,  94 }
\definecolor{secondary}     {RGB}{   0, 105, 180 }
\definecolor{gray}          {RGB}{ 114, 119, 119 }
\definecolor{tertiary}      {RGB}{   0, 159, 227 }

\definecolor{green1}        {RGB}{ 148, 195,  86 }
\definecolor{teal1}         {RGB}{ 138, 203, 193 }
\definecolor{blue1}         {RGB}{ 132, 207, 237 }
\definecolor{magenta1}      {RGB}{ 238, 123, 174 }
\definecolor{red1}          {RGB}{ 240, 130,  98 }
\definecolor{orange1}       {RGB}{ 247, 169,  65 }

\definecolor{green2}        {RGB}{ 101, 179,  46 }
\definecolor{teal2}         {RGB}{   0, 172, 169 }
\definecolor{blue2}         {RGB}{   0, 161, 217 }
\definecolor{magenta2}      {RGB}{ 224,  49, 138 }
\definecolor{red2}          {RGB}{ 232,  65,  44 }
\definecolor{orange2}       {RGB}{ 239, 125,   0 }

\definecolor{green3}        {RGB}{   0, 137,  58 }
\definecolor{teal3}         {RGB}{   0, 131, 141 }
\definecolor{blue3}         {RGB}{   0, 119, 174 }
\definecolor{magenta3}      {RGB}{ 206,   0, 117 }
\definecolor{red3}          {RGB}{ 205,  23,  25 }
\definecolor{orange3}       {RGB}{ 201,  75,  23 }

\definecolor{c11}{RGB}{  0, 173, 239}
\definecolor{c12}{RGB}{  0, 185, 241}
\definecolor{c13}{RGB}{ 65, 199, 244}

\definecolor{c21}{RGB}{  0, 113, 187} 
\definecolor{c22}{RGB}{ 30, 131, 197}
\definecolor{c23}{RGB}{101, 154, 209}

\definecolor{c31}{RGB}{ 87,  63, 152} 
\definecolor{c32}{RGB}{113,  93, 168}
\definecolor{c33}{RGB}{140, 124, 185}

\definecolor{c41}{RGB}{145,  38, 143} 
\definecolor{c42}{RGB}{160,  83, 160}
\definecolor{c43}{RGB}{178, 121, 180}

\definecolor{c51}{RGB}{  0, 140,  78} 
\definecolor{c52}{RGB}{  6, 155, 105}
\definecolor{c53}{RGB}{ 89, 174, 134}

\definecolor{c61}{RGB}{ 97, 187,  69} 
\definecolor{c62}{RGB}{131, 197, 101}
\definecolor{c63}{RGB}{161, 209, 138}

\colorlet{green}  {green2}
\colorlet{teal}   {teal2}
\colorlet{blue}   {blue2}
\colorlet{magenta}{magenta2}
\colorlet{red}    {red2}
\colorlet{orange} {orange2}

\def\setallcolors{%
\gdef\kwcolor{\color{blue}\bfseries}%
\gdef\idcolor{\color{black}}%
\gdef\commentcolor{\color{gray}}%
\gdef\stringcolor{\color{green3}}
\gdef\namespacecolor{\color{c31}}%
\gdef\classcolor{\color{teal3}}%
\gdef\methodcolor{\color{blue}}%
\gdef\macrocolor{\color{c41}}%
\gdef\numcolor{\color{teal3}}%
}
\def\setallgray{%
\gdef\kwcolor{\color{gray}\bfseries}%
\gdef\idcolor{\color{gray}}%
\gdef\commentcolor{\color{gray}}%
\gdef\stringcolor{\color{gray}}%
\gdef\namespacecolor{\color{gray}}%
\gdef\classcolor{\color{gray}}%
\gdef\methodcolor{\color{gray}}%
\gdef\macrocolor{\color{gray}}%
\gdef\numcolor{\color{gray}}%
}
\setallcolors{}
\lstdefinelanguage{coq}{
  morekeywords={Definition, Fixpoint, Lemma, Theorem,
  Example, Proof, Inductive, Notation, Arguments, Hint,
  Instance, Ltac, Class, Variant, CoInductive, CoFixpoint, Notation},
  sensitive=true, 
  morecomment=[s]{(*}{*)}, 
  morestring=[b]{"} 
}
\lstdefinestyle{coq}{
  language=coq,
  emph      = [1]{match, with, end, fun, fix, let, in, assert, if, then, else},
  emphstyle = [1]{\color{blue}},
  emph      = [3]{forall},
  emphstyle = [3]{\color{blue}\rmfamily\itshape},
  emph      = [4]{:, =>, ->},
  emphstyle = [4]{\color{gray}},
  emph      = [5]{},
  emphstyle = [5]{\color{}}
  moredelim=**[is][\classcolor]{@[}{]@},
  moredelim=**[is][\setallgray\color{gray}\aftergroup\setallcolors]{@<}{>@},
  moredelim=**[is][\color{blue}]{@|}{|@},
}
\tikzset{
    inst/.style={rectangle, draw=black!60, fill=white, thick, minimum width=2.5cm, minimum height=0.7cm, align=center, font=\ttfamily\small},
    fusedinst/.style={rectangle, draw=green!60!black, fill=green!10, thick, minimum width=2.5cm, minimum height=1.2cm, align=center, font=\bfseries\small},
    slot/.style={rectangle, draw=black!80, fill=gray!20, thick, minimum width=2.8cm, minimum height=0.8cm, align=center, font=\sffamily\footnotesize},
    fusedslot/.style={rectangle, draw=green!60!black, fill=green!20, thick, minimum width=2.8cm, minimum height=1.6cm, align=center, font=\sffamily\bfseries\small},
    arrow/.style={-{Latex[length=3mm]}, thick, gray},
    fusionarrow/.style={-{Latex[length=3mm]}, ultra thick, green!60!black},
    cross/.style={cross out, draw=red, minimum size=2*(#1-\pgflinewidth), inner sep=0pt, outer sep=0pt, thick},
}

\newcommand{\inst}[1]{\texttt{#1}}

\title{Interaction Tree Semantics for RISC-V: Bridging Compiler and Hardware Verification} 

\titlerunning{Interaction Tree Semantics for RISC-V}

\author{Shuanglong Kan}{Barkhausen Institut, Dresden, Germany}{shuanglong.kan@barkhauseninstitut.org}{}{}

\author{Sebastian Ertel}{Barkhausen Institut, Dresden, Germany}{sebastian.ertel@barkhauseninstitut.org}{}{}

\authorrunning{S. Kan and S. Ertel}

\Copyright{Shuanglong Kan and Sebastian Ertel}



\category{} 

\relatedversion{} 

\acknowledgements{This work was funded by the Agentur f\"ur Innovation in der Cybersicherheit GmbH (Cyberagentur).}

\nolinenumbers 

\EventEditors{John Q. Open and Joan R. Access}
\EventNoEds{2}
\EventLongTitle{42nd Conference on Very Important Topics (CVIT 2016)}
\EventShortTitle{CVIT 2016}
\EventAcronym{CVIT}
\EventYear{2016}
\EventDate{December 24--27, 2016}
\EventLocation{Little Whinging, United Kingdom}
\EventLogo{}
\SeriesVolume{42}
\ArticleNo{23}

\usepackage{ifthen}

\newboolean{showcomments}
\setboolean{showcomments}{false}

\makeatletter
\newcommand{\mynote}[3]{%
  \ifthenelse{\boolean{showcomments}}{%
   \fbox{\bfseries\sffamily\scriptsize#1}%
   {\small\textsf{\emph{\color{#3}{#2}}}}}%
  {%
   \@bsphack
   \@esphack
  }%
}
\makeatother

\definecolor{asparagus}{rgb}{0.53, 0.66, 0.42}

\newcommand{\out}[1]{}

\setboolean{showcomments}{false}

\usepackage{xspace}

\newcommand{\rocq}{\text{Rocq}\xspace}

\newcommand{\riscv}{\mbox{RISC-V}\xspace}

\newcommand{\itree}{\text{ITree}\xspace}
\newcommand{\itrees}{\text{ITrees}\xspace}

\newcommand{\koika}{\text{K\^oika}\xspace}

\setboolean{showcomments}{true}

\begin{document}

\maketitle

\begin{abstract}

The Instruction Set Architecture (ISA) is the contract between
compilers and processors; proving this contract formally demands cross-level
connection to existing mechanized compilers and hardware
implementations.
As an open, modular ISA gaining adoption across embedded, mobile,
and cloud platforms, \riscv makes a formally verified ISA
specification particularly valuable.
However, existing formal \riscv specifications focus on hardware
tooling rather than cross-level verification:
they provide no machine-checked instruction-level properties and lack
support for verifying this contract across levels.
We address these limitations with a formal semantics of the \riscv
ISA in \rocq, built on Interaction Trees (\itrees).
By leveraging \itree bisimulation and refinement, our semantics
enables cross-level verification from compiler IR to hardware
within a single framework.
Our formalization covers a wide spectrum of \riscv extensions.
The correctness of individual instruction semantics is backed by
machine-checked lemmas in \rocq.
We further validate it by extracting an executable simulator that
passes all standard \riscv test suites.
Three case studies demonstrate the effectiveness of our
semantics for cross-level verification:
first, we prove semantic equivalence via bisimulation between
LLVM IR and \riscv code on an array access pattern via
Vellvm (LLVM \itree semantics);
second, we apply translation validation to a specific instruction
reordering for macro-operation fusion, distinguishing safe
reorderings from those that break program-counter-relative addressing;
third, we prove that a \koika hardware ALU correctly implements
all R-type integer operations (e.g., ADD, SUB, AND) against our ISA contract.

\end{abstract}

\section{Introduction}
\label{sec:intro}


%
The Instruction Set Architecture (ISA) is arguably the most important
interface in a computer system~\cite{Waterman2016Design}.
It defines the contract between software and hardware:
compilers rely on the ISA to generate correct code, and
processor designers implement it in silicon.
\riscv{}~\cite{Asanovic2014Free,RISCVUnpriv2019,RISCVPriv2021} has emerged
as a prominent open standard ISA with a rich ecosystem of
compilers (GCC, LLVM) and hardware designs
(BOOM~\cite{Celio2015BOOM}, Rocket~\cite{Asanovic2016RocketChip}).
To bring formal verification to this ecosystem, a machine-checked ISA
semantics is needed as the bridge that enables
verifying both sides independently---compiler correctness on one side,
hardware refinement on the other---while
guaranteeing end-to-end correctness of the whole system.
%


%

Due to this pivotal role, the correctness of the \riscv ISA
semantics itself is critical for the soundness of all downstream
verification efforts.
Moreover, the formal semantics must serve as a shared foundation
for multiple verification tasks along the verification chain from
compiler to hardware.
First, \emph{compiler correctness} must show that compiled
code preserves the semantics of the source
program~\cite{Leroy2009CACM,Leroy2006POPL,Kumar2014CakeML,Tan2016CakeMLBackend}
with respect to the \riscv ISA specification.
Second, \emph{microarchitectural optimizations} such as
macro-operation fusion---which combines adjacent instructions into
single operations~\cite{Shen2024MacroFusion}---may require the
compiler to reorder instructions to create fusion opportunities;
proving that such reorderings preserve program semantics demands
a formal ISA specification shared by both compiler and hardware.
Third, \emph{hardware refinement} must show that a processor
implementation faithfully realizes the \riscv ISA specification,
so that different microarchitectures can be verified against the
same formal contract.
To support these diverse tasks, the semantics must interface with
different software and hardware languages across abstraction
levels.
%
%

The Sail language~\cite{Armstrong2019Sail} provides the official
  \riscv{} reference specification, adopted by RISC-V International,
  with support for generating theorem prover definitions for
  Isabelle, HOL4, and \rocq.
  The MIT PLV group derives \rocq specifications from Haskell
  semantics~\cite{bourgeat2021multipurpose,Bourgeat2023RiscvMonads},
  mainly serving the Bedrock2 verified systems stack.
  In both cases, the generated \rocq definitions lack bisimulation
  relations for establishing semantic equivalence across abstraction
  levels, and are difficult to integrate with independently developed
  compiler semantics (e.g., Vellvm~\cite{Zakowski2021LLVM} for LLVM IR).
  Since the semantics are defined externally,
  neither approach provides machine-checked proofs within \rocq
  that individual instructions satisfy their intended properties.
  Moreover, the translation to \rocq is itself not formally verified:
  the semantic gap between the source languages (Sail, Haskell) and
  \rocq makes the translation error-prone, placing it in the trusted
  computing base of any downstream verification effort.
  %
%

We address these limitations by defining \riscv semantics
natively in \rocq using Interaction Trees
(\itrees)~\cite{Xia2020ITrees}, eliminating unverified translations
and ensuring that all proofs are machine-checked within a single
trusted framework.
\itrees are a coinductive data structure for representing effectful and
potentially nonterminating
computations~\cite{He2020GPaco,Park1981Bisimulation,Milner1989CCS,Sangiorgi2009Origins}
that support compositional reasoning as well as bisimulation and
refinement---properties we exploit for cross-level verification.

First, because our semantics shares the \itree foundation with
existing \itree language semantics---including
LLVM IR~\cite{Zakowski2021LLVM},
Jasmin~\cite{ArranzOlmos2025Jasmin},
PureCake~\cite{Kanabar2023PureCake} (adapted for HOL4),
and concurrent memory models~\cite{Chappe2025MonadicInterpreters}---we
can prove semantic equivalence via bisimulation between them.
We demonstrate this with Vellvm (LLVM IR
semantics)---a key step toward end-to-end
compiler verification.
The \itree ecosystem also integrates with the Iris separation logic
framework~\cite{Jung2018Iris} via Guarded Interaction
Trees~\cite{Frumin2024GITrees} and modular program
logics~\cite{Vistrup2025ProgramLogics}, opening a path in the future to build a program
logic for \riscv without starting from scratch and to connect memory safety
properties established at the language level to memory behavior in our
\riscv semantics.
Second, we leverage \itree refinement to verify hardware
implementations against our ISA semantics.
While \itree refinement has been applied in the software
domain---for example, Koh et~al.~\cite{Koh2019DeepWeb} prove that a C
network server implementation refines its \itree linear specification---to the best of our
knowledge, our work is the first to apply
\itree-based refinement to hardware verification, where the
\riscv ISA semantics serves as the specification that processor
implementations must refine.
%
%

%



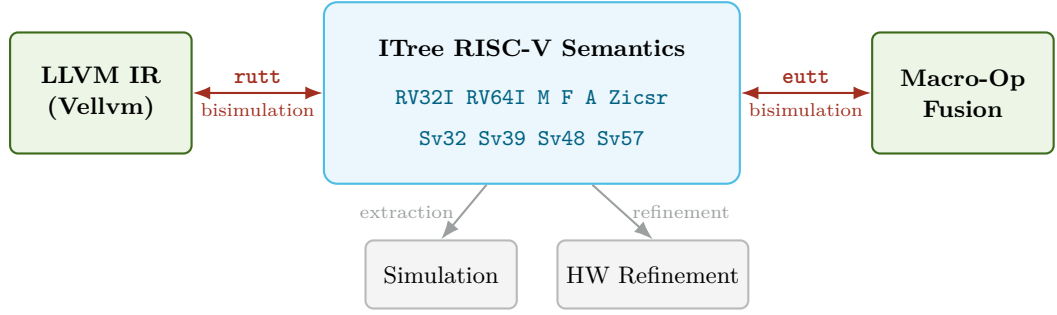
\begin{figure}[t]
    \centering
        \begin{tikzpicture}[
        core/.style={rectangle, draw=blue!70, fill=blue!8, thick,
            minimum width=5.5cm, minimum height=2.4cm, align=center,
            font=\small\bfseries, rounded corners=5pt},
        usecase/.style={rectangle, draw=green!60!black, fill=green!12, thick,
            minimum width=2.4cm, minimum height=1.6cm, align=center,
            font=\small\bfseries, rounded corners=3pt},
        secondary/.style={rectangle, draw=gray!50, fill=gray!8, thick,
            minimum width=2cm, minimum height=0.9cm, align=center,
            font=\small, rounded corners=3pt},
        future/.style={rectangle, draw=gray!50, fill=gray!8, thick,
            minimum width=2cm, minimum height=0.9cm, align=center,
            font=\small, rounded corners=3pt},
        bisim/.style={Latex-Latex, thick, red!70!black},
        downarrow/.style={-{Latex[length=2.5mm]}, thick, gray!70},
        component/.style={font=\footnotesize\ttfamily, text=blue!60!black},
    ]
        \node[core] (itree) at (0, 0) {};
        \node[font=\small\bfseries] at (0, 0.6) {ITree RISC-V Semantics};
        \node[component] at (0, -0.05) {RV32I\enspace RV64I\enspace M\enspace F\enspace A\enspace Zicsr};
        \node[component] at (0, -0.6) {Sv32\enspace Sv39\enspace Sv48\enspace Sv57};

        \node[usecase] (llvm) at (-5.7, 0) {LLVM IR\\(Vellvm)};

        \node[usecase] (fusion) at (5.7, 0) {Macro-Op\\Fusion};

        \draw[bisim] (llvm.east) -- (itree.west)
            node[midway, above, font=\footnotesize\bfseries] {\texttt{rutt}}
            node[midway, below, font=\scriptsize] {bisimulation};
        \draw[bisim] (itree.east) -- (fusion.west)
            node[midway, above, font=\footnotesize\bfseries] {\texttt{eutt}}
            node[midway, below, font=\scriptsize] {bisimulation};

        \node[secondary] (sim) at (-1.2, -2.4) {Simulation};
        \node[secondary] (hw) at (1.6, -2.4) {HW Refinement};

        \draw[downarrow] ([xshift=-0.6cm]itree.south) -- (sim.north)
            node[midway, left, font=\scriptsize] {extraction};
        \draw[downarrow] ([xshift=0.8cm]itree.south) -- (hw.north)
            node[midway, right, font=\scriptsize] {refinement};

    \end{tikzpicture}
    \caption{Architecture of the ITree-based RISC-V formalization.}
    \label{fig:architecture}
\end{figure}

%
Figure~\ref{fig:architecture} illustrates the architecture of our
\itree{}-based \riscv semantics.
The formalization covers both unprivileged and privileged aspects of
the \riscv ISA and closely follows the authoritative RISC-V standard
Sail specification, which also facilitates proving consistency between
the two semantics in the future.

\subparagraph*{Contributions.}
Our key contribution is a machine-checked \riscv semantics that,
for the first time, enables cross-level verification from compiler
IR to hardware within a single \itree-based framework. In detail:
\begin{description}
  \item[Machine-checked \riscv semantics.]
        We present an \itree{}-based formal semantics of \riscv in
        \rocq, covering the RV32/64 base integer instructions (I),
        integer multiplication and division (M),
        single-precision floating-point (F),
        atomic instructions (A),
        control and status registers (Zicsr),
        and the Sv32/Sv39/Sv48/Sv57 virtual memory
        systems~\cite{RISCVPriv2021}.
        The correctness of individual instruction semantics is backed
        by 131 machine-checked lemmas in \rocq.
        These lemmas required developing proof patterns for reasoning
        about bitvector operations, memory effects, and exception behavior
        within the \itree{} framework—techniques that generalize beyond
        \riscv{} to other ISA formalizations.
        We further validate the semantics by extracting an executable \riscv
        simulator in OCaml and running the official \riscv test suites,
        which proved essential for catching subtle human misinterpretations
        of the specification.
  \item[Case studies.]
        Three case studies demonstrate our semantics serves as
        a shared foundation for verifying the \riscv ISA contract across abstraction levels:
        (1)~semantic equivalence via bisimulation between LLVM IR
        and \riscv code on an array access pattern via
        Vellvm~\cite{Zakowski2021LLVM};
        (2)~translation validation of a specific instruction
        reordering for macro-operation fusion, distinguishing safe
        reorderings from those that break Program Counter (PC)-relative addressing; and
        (3)~refinement of a
        \textsc{\koika}~\cite{Bourgeat2020Koika} hardware ALU,
        proving it correctly implements all R-type integer operations
        (e.g., ADD, SUB, AND) against our ISA contract.
\end{description}

\subparagraph*{Organization.}
Section~\ref{sec:itrees} provides background on Interaction Trees.
We then present our \itree{}-based \riscv formalization in Section~\ref{sec:itree-semantics},
validate it by simulation against the official test suite in Section~\ref{sec:validation},
and demonstrate its use through case studies in Section~\ref{sec:evaluation}.
We discuss related work in Section~\ref{sec:related} and
conclude in Section~\ref{sec:conclusion}.

\section{Interaction Trees}
\label{sec:itrees}

Interaction Trees (ITrees)~\cite{Xia2020ITrees} are a coinductive data structure for representing effectful and potentially nonterminating computations in type theory.
They provide a principled way to embed impure programs into a pure logical framework like Rocq, enabling formal reasoning about programs with side effects, I/O, and infinite behavior.

\subsection{Definition and Structure}

An Interaction Tree is parameterized by an \emph{event type} \texttt{E : Type -> Type} that describes the interface to the external world, and a \emph{return type} \texttt{R} for the computation's final result.
The event type is a type constructor: given a response type \texttt{X}, the type \texttt{E X} represents events that, when handled by the environment, produce a response value of type \texttt{X}.
For example, a register read event \texttt{ReadReg r: E Word} returns a word, while a register write \texttt{WriteReg r v: E unit} returns unit.
The coinductive definition has three constructors (Listing~\ref{lst:itree-def}):

\noindent\begin{minipage}{\linewidth}
\begin{lstlisting}[language=coq, caption={Interaction Tree definition}, label={lst:itree-def}]
CoInductive itree (E : Type -> Type) (R : Type) : Type :=
| Ret (r : R)                              (* Return a value *)
| Tau (t : itree E R)                      (* Silent step *)
| Vis (e : E X) (k : X -> itree E R)       (* Visible event *)
\end{lstlisting}
\end{minipage}

The \texttt{Ret r} constructor indicates the computation has terminated with value \texttt{r}.
The \texttt{Tau t} constructor represents a silent internal step (written $\tau$), continuing as \texttt{t}; this enables representation of divergent computations and internal reduction steps.
The \texttt{Vis e k} constructor emits a visible event \texttt{e} of type \texttt{E X} for some response type \texttt{X}, then applies the continuation \texttt{k} to the environment's response.

The coinductive nature of ITrees allows infinite unfoldings, making them suitable for modeling nonterminating programs such as operating systems, servers, and hardware execution loops.
The \texttt{CoInductive} definition is shown for presentation; in practice, the library defines equivalence relations via the Paco library~\cite{He2020GPaco} (a greatest fixed point combinator) and provides equational lemmas (e.g., monad laws, unfolding rules) so that users reason by rewriting rather than writing coinductive proofs directly.

\subsection{Monadic Interface}

For every event type \texttt{E}, \texttt{itree E} forms a monad, enabling sequential composition of effectful computations (Listing~\ref{lst:itree-monad}).
The \texttt{bind} operation (written \texttt{x <- t;; k}) threads the result of computation \texttt{t} into continuation \texttt{k}, while \texttt{trigger} emits an event and returns its response.
Listing~\ref{lst:itree-example} shows a register swap example using these operations.

\noindent
\begin{minipage}[t]{0.48\textwidth}
\begin{lstlisting}[language=coq, caption={Monadic operations}, label={lst:itree-monad}, numbers=none]
Definition ret {E R} (r: R): itree E R
  := Ret r.

CoFixpoint bind {E R S}
  (t: itree E R) (k: R -> itree E S)
  : itree E S :=
  match t with
  | Ret r => k r
  | Tau t' => Tau (bind t' k)
  | Vis e k' => Vis e (fun x =>
                  bind (k' x) k) end.
\end{lstlisting}
\end{minipage}
\hfill
\begin{minipage}[t]{0.48\textwidth}
\begin{lstlisting}[language=coq, caption={Example: register swap}, label={lst:itree-example}, numbers=none]
Notation "x <- t ;; k"
  := (bind t (fun x => k)).

Definition trigger {E X} (e: E X)
  : itree E X := Vis e (fun x => Ret x).

Definition swap r1 r2: itree RegE unit :=
  v1 <- trigger (ReadReg r1) ;;
  v2 <- trigger (ReadReg r2) ;;
  trigger (WriteReg r1 v2) ;;
  trigger (WriteReg r2 v1).
\end{lstlisting}
\end{minipage}

\vspace{0.5em}
\noindent
The \texttt{swap} function reads both registers via \texttt{ReadReg} events (of type \texttt{RegE Word}, returning the register value), then writes each value to the other register via \texttt{WriteReg} events (of type \texttt{RegE unit}).
Each \texttt{trigger} call emits an event and binds the environment's response to the continuation.

\subsection{Event Handlers and Interpretation}

A crucial feature of ITrees is the separation between \emph{specifying} effects (via events) and \emph{implementing} them (via handlers).
An event handler interprets events from one signature into computations over another (Listing~\ref{lst:handler}):

\noindent\begin{minipage}{\linewidth}
\begin{lstlisting}[language=coq, caption={Event handler, interpretation, and example}, label={lst:handler}]
Definition handler (E F : Type -> Type) : Type := forall X, E X -> itree F X.
Definition interp {E F} (h : handler E F) : forall R, itree E R -> itree F R.

Definition reg_handler : handler RegE StateE :=
  fun X e =>
    match e with
    | ReadReg r  => trigger (Get r)
    | WriteReg r v => trigger (Put r v)
    end.
\end{lstlisting}
\end{minipage}

The \texttt{reg\_handler} example translates abstract \texttt{RegE} events into \texttt{StateE} operations: \texttt{ReadReg} becomes a \texttt{Get} that retrieves a value from the register file state, and \texttt{WriteReg} becomes a \texttt{Put} that updates it.

The \texttt{interp} function applies a handler to an ITree, recursively replacing each \texttt{Vis} node's event with the handler's implementation.
For example, \texttt{interp reg\_handler (swap r1 r2)} produces an ITree over \texttt{StateE} events.
A second interpretation step can then handle \texttt{StateE} by threading an explicit register file through the computation, so that starting from \texttt{\{r1 $\mapsto$ 5, r2 $\mapsto$ 7\}} yields final state \texttt{\{r1 $\mapsto$ 7, r2 $\mapsto$ 5\}}.

\subsection{Equivalence Relations}

The ITree library provides a family of equivalence relations.
The base relation \texttt{eqit b1 b2} is parameterized by two booleans controlling whether \texttt{Tau} steps can be stripped on each side.
Instantiations include \texttt{eq\_itree} (strong bisimulation, no tau stripping), \texttt{eutt} (equivalence up to taus, stripping on both sides), and \texttt{euttge} (refinement, stripping taus only on the left).
\texttt{rutt} (relational up-to-taus) generalizes \texttt{eutt} to heterogeneous events.
We introduce \texttt{eutt} and \texttt{rutt} below, as these are the two main relations we use in our verification.

\subparagraph*{Weak Bisimulation (\texttt{eutt}).}
\emph{Equivalence up to taus} (\texttt{eutt}) equates ITrees with the same event type that differ only in the number of internal \texttt{Tau} steps (Listing~\ref{lst:eutt}):

\noindent\begin{minipage}{\linewidth}
\begin{lstlisting}[language=coq, caption={Homogeneous weak bisimulation}, label={lst:eutt}]
eutt (RR : R1 -> R2 -> Prop) : itree E R1 -> itree E R2 -> Prop
\end{lstlisting}
\end{minipage}

Given two \itrees \texttt{t1}~and~\texttt{t2}, the relation
\texttt{eutt RR t1 t2} holds when internal \texttt{Tau}
steps on either side are ignored, when both trees emit the same
event, their continuations are related by \texttt{eutt} for all
responses, and when both trees terminate, their final return values are related by \texttt{RR}.
The \texttt{eutt} relation is an equivalence and a congruence with respect to \texttt{bind}, enabling equational reasoning.

\subparagraph*{Heterogeneous Weak Bisimulation (\texttt{rutt}).}
For connecting semantics at different abstraction levels (e.g., LLVM IR to RISC-V), ITree provides \emph{relational up-to-tau} (\texttt{rutt}), a heterogeneous weak bisimulation that relates ITrees with different event types through user-specified relations (Listing~\ref{lst:rutt}):

\noindent\begin{minipage}{\linewidth}
\begin{lstlisting}[language=coq, caption={Heterogeneous weak bisimulation}, label={lst:rutt}]
rutt (REv : forall A B, E1 A -> E2 B -> Prop)
     (RAns : forall A B, E1 A -> A -> E2 B -> B -> Prop)
     (RR : R1 -> R2 -> Prop) : itree E1 R1 -> itree E2 R2 -> Prop
\end{lstlisting}
\end{minipage}

Given two \itrees \texttt{t1:itree E1 R1} and
\texttt{t2:itree E2 R2}, the relation
\texttt{rutt REv RAns RR t1 t2} is parameterized by three
relations:
\texttt{REv} (the \emph{event relation}) specifies which events
from \texttt{t1} and \texttt{t2} correspond to each other;
\texttt{RAns} (the \emph{response relation}) specifies how the
environment's responses to matched events must correspond; and
\texttt{RR} (the \emph{return relation}) specifies how the final
return values must correspond.
The relation holds when internal \texttt{Tau} steps on either side
are ignored, every event \texttt{e1} emitted by \texttt{t1} is
matched by a \texttt{REv}-related event \texttt{e2} from
\texttt{t2}, the continuations are related by \texttt{rutt} for any
pair of \texttt{RAns}-related responses, and the final return values
are related by \texttt{RR}.

\section{ITree-Based RISC-V Semantics}
\label{sec:itree-semantics}

\riscv defines binary instructions that perform computation, access registers, interact with memory, and transfer control flow.
A processor executes them in a fetch-decode-execute cycle.
Our \itree{}-based formalization models both the cycle and the per-instruction semantics in the execution stage of the fetch-decode-execute cycle: each instruction is an \itree that emits events capturing its interactions with registers and memory.
In the following, we describe the events that instructions emit (\S\ref{subsec:events}), the per-instruction semantics (\S\ref{subsec:instructions}), and correctness proofs for individual instructions (\S\ref{subsec:correctness}).

\subsection{Event Hierarchy}
\label{subsec:events}

Events are primitives for accessing registers and memory.
Instruction semantics are composed of these events, which are then interpreted by handlers.
We classify events into register access events and memory access events.

\subparagraph*{Register Access Events.}
Listing~\ref{lst:processore} shows the register access event type, covering integer registers (\texttt{x0}--\texttt{x31}, \texttt{RegRead} and \texttt{RegWrite}), floating-point registers (\texttt{f0}--\texttt{f31}, \texttt{FPRegRead} and \texttt{FPRegWrite}), the program counter (\texttt{pc}, \texttt{PCRead} and \texttt{PCWrite}), and Control and Status Registers (CSRs, \texttt{CSRRead} and \texttt{CSRWrite}).
The type \texttt{bv n} represents an $n$-bit bitvector, \texttt{XLEN} is the native word width of the architecture (32 for RV32, 64 for RV64), and \texttt{regidx} is a register index of type \texttt{fin 32} (a finite type with 32 elements).
Events are parameterized by their return type: read operations return values while write operations return \texttt{unit}.
CSR operations return \texttt{Result} because accessing invalid or privileged CSRs can fail.

\noindent\begin{minipage}{\linewidth}
\begin{lstlisting}[language=coq, caption={Processor event type}, label={lst:processore}]
Variant ProcessorE : Type -> Type :=
  | RegRead (r : regidx) : ProcessorE (bv XLEN)
  | RegWrite (r : regidx) (d : bv XLEN) : ProcessorE unit
  (* F extension: floating-point register operations *)
  | FPRegRead (r : regidx) : ProcessorE (bv 32)
  | FPRegWrite (r : regidx) (d : bv 32) : ProcessorE unit
  | PCRead : ProcessorE (bv XLEN)
  | PCWrite (new_pc : bv XLEN) : ProcessorE unit
  (* Zicsr extension: CSR operations *)
  | CSRRead (addr : bv 12) : ProcessorE (Result (bv XLEN))
  | CSRWrite (addr : bv 12) (val : bv XLEN) : ProcessorE (Result unit).
\end{lstlisting}
\end{minipage}

\subparagraph*{Memory Access Events.}
We define two levels of memory access: virtual memory events (\texttt{VMemE}) and physical memory events (\texttt{PMemE}), as shown in Listing~\ref{lst:vmeme}.
This separation enables reasoning about instruction semantics independently of address translation---proofs at the virtual memory level need not consider page table walks, while translation correctness can be established separately.
RISC-V instruction semantics use only \texttt{VMemE}; when reasoning about physical memory is required, a handler interprets virtual memory events into physical memory events.
Virtual addresses are split into a base address (\texttt{vaddr}) and an offset (\texttt{offset}), following the RISC-V load/store addressing mode.
The \texttt{res} parameter indicates whether the access is a reservation for the A extension's load-reserved/store-conditional (\inst{LR}/\inst{SC}) instructions, which implement atomic read-modify-write sequences; most instructions set this to \texttt{false}.
Despite the name, \texttt{VMemE} also supports direct physical memory access: when the \texttt{satp} register is in Bare mode, address translation is bypassed and addresses are used directly as physical addresses.

\noindent\begin{minipage}{\linewidth}
\begin{lstlisting}[language=coq, caption={Memory event types}, label={lst:vmeme}]
Variant VMemE : Type -> Type :=
  | VMemRead (vaddr : bv XLEN) (offset : bv XLEN)
             (width : WIDTH) (res : bool) : VMemE (Result (bv XLEN))
  | VMemWrite (vaddr : bv XLEN) (offset : bv XLEN)
              (width : WIDTH) (data : bv XLEN) (res : bool) : VMemE (Result unit)
  | VMemInstrFetch (addr : bv XLEN) : VMemE (Result (bv 32)).

Variant PMemE (paddr_width : N) : Type -> Type :=
  | PMemRead (paddr : bv paddr_width) (width : WIDTH)
               : PMemE paddr_width (Result (bv width))
  | PMemWrite (paddr : bv paddr_width) (width : WIDTH)
              (data : bv  width) : PMemE paddr_width (Result unit).
\end{lstlisting}
\end{minipage}

\noindent\begin{minipage}{\linewidth}
\begin{lstlisting}[language=coq, caption={Instruction variant type}, label={lst:instr}]
Variant instr (regidx : Type) :=
  (* I extension: base integer instructions *)
  | ITYPE (imm : bv 12) (rs1 rd : regidx) (op : itype_op_type)
  | RTYPE (rs1 rs2 rd : regidx) (op : rtype_op_type)
  | BTYPE (imm : bv 13) (rs1 rs2 : regidx) (op : btype_op_type)
  | ...
  | JALTYPE (imm : bv 21) (rd : regidx)
  | JALRTYPE (imm : bv 12) (rs1 rd : regidx)
  | LOAD (imm : bv 12) (rs1 rd : regidx) (is_unsigned : bool) (width : WIDTH)
  | STORE (imm : bv 12) (rs1 rs2 : regidx) (width : WIDTH)
  | FENCE (pred succ : bv 4)
  (* RV64I: word-width variants *)
  | ADDIWOP (imm : bv 12) (rs1 rd : regidx)
  | RTYPEW (rs1 rs2 rd : regidx) (op : rtypew_op_type)
  (* M extension: multiplication and division *)
  | MTYPE (rs1 rs2 rd : regidx) (op : mtype_op_type)
  | MTYPEW (rs1 rs2 rd : regidx) (op : mtypew_op_type)
  (* A extension: atomic memory operations *)
  | ATYPE (rs1 rs2 rd : regidx) (width : WIDTH) (aq rl : bool) (op : amo_op_type)
  (* Zicsr extension: CSR instructions *)
  | CSRTYPE (csr : bv 12) (rs1 rd : regidx) (op : csrtype_op_type)
  | CSRITYPE (csr : bv 12) (uimm : bv 5) (rd : regidx) (op : csritype_op_type)
  (* F extension: floating-point operations *)
  | FLOAD (imm : bv 12) (rs1 fd : regidx)
  | FARITH (fs1 fs2 fd : regidx) (rm : bv 3) (op : ftype_arith_op)
  | ...
  | ILLEGAL.
\end{lstlisting}
\end{minipage}

\subsection{Instruction Semantics}
\label{subsec:instructions}

Following the official Sail RISC-V specification~\cite{Armstrong2019Sail}, we define instructions as a variant type where each constructor corresponds to a RISC-V instruction format (Listing~\ref{lst:instr}).
For example, \texttt{RTYPE} groups register-register arithmetic operations, e.g., the assembly instruction \texttt{add x1, x2, x3} (which adds registers \texttt{x2} and \texttt{x3} into \texttt{x1}) is represented as \texttt{RTYPE x2 x3 x1 ADD}.
An operation type parameter (e.g., \texttt{itype\_op\_type}) distinguishes individual instructions within each format.
The type is parameterized by \texttt{regidx}, enabling shared definitions across RV32 and RV64.

\smallskip\noindent\textbf{Integer and Load Instructions.}
Listing~\ref{lst:rtype} shows representative instruction semantics.
The R-type arithmetic semantics follows the RISC-V specification directly: read the two source registers \texttt{rs1} and \texttt{rs2}, compute the result based on the operation type, and write to the destination register \texttt{rd}.
Load instructions additionally compute the effective address from \texttt{rs1} and a sign-extended immediate, then emit a \texttt{VMemRead} event.
On success, the loaded data is extended and written to \texttt{rd}; on failure, the memory exception is propagated.

\noindent\begin{minipage}{\linewidth}
\begin{lstlisting}[language=coq, caption={R-type and load instruction execution}, label={lst:rtype}]
Definition exec_RTYPE rs1 rs2 rd op : itree RISCV_Event ExecResult :=
  x_rs1 <- trigger (RegRead rs1);;
  x_rs2 <- trigger (RegRead rs2);;
  let result := match op with
    | ADD => bv_add x_rs1 x_rs2
    | SUB => bv_sub x_rs1 x_rs2
    | SLT => zero_extend XLEN (bv_slt x_rs1 x_rs2) | ... end in
  trigger (RegWrite rd result);; ret Retire_Success.

Definition exec_LOAD imm rs1 rd is_unsigned width : itree RISCV_Event ExecResult :=
  vaddr <- trigger (RegRead rs1);;
  let offset := sign_extend XLEN imm in
  mem_result <- trigger (VMemRead vaddr offset width false);;
  match mem_result with
  | Ok data =>
      let ext := bv_extract 0 (width_to_bits width) data in
      let result := if is_unsigned then data
                    else sign_extend XLEN ext in
      trigger (RegWrite rd result);; ret Retire_Success
  | Err e => ret e
  end.
\end{lstlisting}
\end{minipage}

\smallskip\noindent\textbf{Floating-Point Fused Multiply-Add (FMA).}
The fused multiply-add family (FMADD.S, FMSUB.S, FNMADD.S, FNMSUB.S)
illustrates a key modeling challenge: \emph{data-dependent side effects}.
Unlike integer instructions whose event traces are fixed, FMA (Listing~\ref{lst:ffma}) interacts with CSR registers in ways that depend on runtime state, creating two sources of control-flow variability:
\emph{(i)}~the rounding mode field (\texttt{rm\_bits}) can be \texttt{DYN} (dynamic), meaning the effective rounding mode must be read at runtime from the floating-point rounding mode CSR (\texttt{frm}) via \texttt{get\_rounding\_mode};
\emph{(ii)}~IEEE~754 exception flags trigger \texttt{update\_fflags} to conditionally read-modify-write the \texttt{fflags} CSR only when at least one flag is raised.
Together, these mean a single FMA instruction can emit between 4 and 7 events.

\noindent\begin{minipage}{\linewidth}
\begin{lstlisting}[language=coq, caption={ITree semantics for the FMA instruction family.}, label={lst:ffma}]
Definition exec_FFMA (fs1 fs2 fs3 fd : regidx) (rm_bits : bv 3) (op : ftype_fma_op)
  : itree (ProcessorE +' VMemE) ExecResult :=
  rm <- get_rounding_mode rm_bits;; a <- trigger (FPRegRead fs1);;
  b <- trigger (FPRegRead fs2);; c <- trigger (FPRegRead fs3);;
  let '(result, flags) := match op with
    | FMADD_S  => fmadd_s  a b c rm | FMSUB_S  => fmsub_s  a b c rm
    | FNMADD_S => fnmadd_s a b c rm | FNMSUB_S => fnmsub_s a b c rm
    end in trigger (FPRegWrite fd result);; update_fflags flags;; Ret Retire_Success.
\end{lstlisting}
\end{minipage}

\noindent
The function takes three source registers \texttt{fs1}, \texttt{fs2}, \texttt{fs3}, a destination register \texttt{fd}, a rounding mode encoding \texttt{rm\_bits}, and an operation selector \texttt{op}.
It reads the three floating-point operands, dispatches to the corresponding pure computation (\texttt{fmadd\_s} for $a \times b + c$, \texttt{fmsub\_s} for $a \times b - c$, \texttt{fnmadd\_s} for $-(a \times b + c)$, \texttt{fnmsub\_s} for $-(a \times b - c)$), writes the result to \texttt{fd}, and accumulates any exception flags via \texttt{update\_fflags}.

\smallskip\noindent\textbf{Page Table Walk.}
The virtual memory subsystem implements the RISC-V Sv32/39/48/57 page table formats.
RISC-V uses a multi-level page table structure: rather than storing one flat mapping from virtual to physical addresses (which would require millions of entries), the translation is split across multiple levels of smaller tables, so only the portions of the address space actually in use need to be allocated.
The number of levels varies by mode---Sv32 uses 2 levels, Sv39 uses 3, Sv48 uses 4, and Sv57 uses 5---trading off address space size against walk depth.
We implement a single configurable page table walk algorithm, parameterized by the number of levels, to support all four modes.

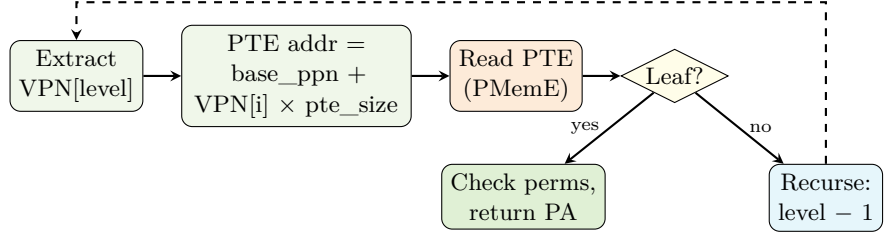
\begin{figure}[t]
\centering
\begin{tikzpicture}[
    box/.style={draw, rounded corners, minimum height=0.7cm, font=\small, align=center},
    decision/.style={draw, diamond, aspect=2, font=\small, inner sep=1pt},
    arr/.style={->, thick, >=stealth},
    node distance=0.6cm
]
    \node[box, fill=green!10] (extract) {Extract\\VPN[level]};
    \node[box, fill=green!10, right=0.5cm of extract, text width=2.8cm] (compute) {PTE addr =\\base\_ppn + VPN[i] $\times$ pte\_size};
    \node[box, fill=orange!15, right=0.5cm of compute] (read) {Read PTE\\(PMemE)};
    \node[decision, fill=yellow!10, right=0.5cm of read] (leaf) {Leaf?};

    \draw[arr] (extract) -- (compute);
    \draw[arr] (compute) -- (read);
    \draw[arr] (read) -- (leaf);

    \node[box, fill=green!20, below=0.8cm of leaf, xshift=-2cm] (result) {Check perms,\\return PA};
    \draw[arr] (leaf) -- node[left, font=\scriptsize]{yes} (result);

    \node[box, fill=blue!10, below=0.8cm of leaf, xshift=2cm] (recurse) {Recurse:\\level $-$ 1};
    \draw[arr] (leaf) -- node[right, font=\scriptsize]{no} (recurse);
    \draw[arr, dashed] (recurse.north) |- ([yshift=0.5cm]extract.north) -- (extract.north);

\end{tikzpicture}
\caption{Page table walk algorithm.}
\label{fig:ptwalk}
\end{figure}

Figure~\ref{fig:ptwalk} illustrates our formalization of the walk algorithm (\texttt{pt\_walk}).
At each level, the algorithm extracts the virtual page number (VPN) bits for that level, computes the physical address of the page table entry (PTE) by indexing into the current page table, and reads the PTE via a \texttt{PMemE} event.
If the PTE is a non-leaf pointer, the algorithm recurses to the next lower level using the physical page number (PPN) from the PTE as the new base address.
If the PTE is a leaf, the walk checks permissions and returns the translated physical address.
The recursion terminates structurally on the level count, ensuring totality in Rocq.

\subsection{Per-Instruction Correctness Proofs}
\label{subsec:correctness}

Because instruction semantics are defined as ITrees over events, we
can prove correctness lemmas by interpreting these trees against a
concrete state.
The \texttt{combined\_handler} is a state-monad handler that
interprets both \texttt{ProcessorE} and \texttt{VMemE} events
against the concrete processor state: \texttt{ProcessorE} events
(e.g., \texttt{RegRead}, \texttt{RegWrite}) are resolved by
reading from or writing to the register file, while \texttt{VMemE}
events are resolved through address translation and memory access.
The function \texttt{interp\_state} applies the
\texttt{combined\_handler} to an \itree,
threading the processor state through each event and producing a
pure \itree with no remaining events.

For example, the following lemma states that
\texttt{exec\_RTYPE}, when interpreted with the
\texttt{combined\_handler}, updates exactly the destination register
with the expected computed value:

\noindent\begin{minipage}{\linewidth}
\begin{lstlisting}[language=coq, caption={Correctness lemma for R-type instructions}, label={lst:rtype_correct}]
Lemma exec_RTYPE_correct : forall s rs1 rs2 rd op,
  rs1 <> x0 -> rs2 <> x0 -> rd <> x0 ->
  let wval := match op with
    | ADD => bv_add | SUB => bv_sub
    | SLT => fun x y =>
        zero_extend XLEN (bv_slt x y) | ... end in
  interp_state combined_handler (exec_RTYPE rs1 rs2 rd op) s
    <|$\approx$|> Ret (set_gp rd (wval (get_gp rs1 s) (get_gp rs2 s)) s, Retire_Success).
\end{lstlisting}
\end{minipage}

\noindent
The preconditions \texttt{rs1 $\neq$ x0}, \texttt{rs2 $\neq$ x0}, and \texttt{rd $\neq$ x0} exclude the hardwired zero register, which always reads as zero.
\texttt{get\_gp} and \texttt{set\_gp} read and write a general-purpose integer register, and \texttt{wval} maps each operation type to its bitvector computation (e.g., \texttt{bv\_add} for \inst{ADD}).
The bisimilarity ($\approx$) asserts that the resulting \itree is
equivalent to a single \texttt{Ret} producing the updated state
with register \texttt{rd} set to the expected value.

Our \rocq development contains 1,258 definitions and functions
(\texttt{Definition}, \texttt{Fixpoint}, \texttt{Inductive}, and
\texttt{Variant}) and 360 lemmas. Of these, 131 are per-instruction correctness
proofs establishing that each modeled instruction produces the expected state update. 27 lemmas verify
virtual memory properties including page table walk correctness and permission
checking. 60 lemmas ensure consistency of instruction encoding and decoding, specifically the lemma $\texttt{decode}(\texttt{encode}(i)) = i$ for all legal instructions.
The remaining 142 serve as integration tests: they execute multi-instruction programs through the full fetch-decode-execute cycle and assert that the final machine state matches expected values.

\section{Semantics Validation by Simulation}
\label{sec:validation}

The correctness proofs in Section~\ref{subsec:correctness} verify properties relative to our formalization, but if the \rocq definitions misinterpret the \riscv specification, the proofs will still hold while the semantics is wrong.
To guard against this, we extract an executable simulator and test it against the official \riscv test suite~\cite{RISCVTests}.
The simulator comprises an OCaml front-end for parsing ELF files and
a back-end extracted from \rocq that executes instructions using our \itree{}-based semantics.

Table~\ref{tab:test-coverage} summarizes our test coverage across RV32 and RV64 architectures.
Each test file is an ELF binary containing multiple test cases for specific instructions or features.
In total, we executed and passed 172 test files comprising 3,228 individual test cases,
covering all official tests for the extensions we support (I, M, F, A, Zicsr).

This testing uncovered several subtle bugs in our initial formalization.
For example, the \texttt{csrr a0, mhartid} test revealed a bug in our Zicsr implementation.
The pseudo-instruction \texttt{csrr rd, csr} assembles to \texttt{csrrs rd, csr, x0}, which reads a CSR and writes the value back.
Our initial implementation performed this CSR write unconditionally, but the RISC-V specification mandates that when \texttt{rs1 = x0}, no CSR write shall occur.
Since \texttt{mhartid} is read-only, our erroneous write attempt triggered an illegal instruction exception, whereas the correct behavior is to perform a read-only access and complete without exception.
Such subtle corner cases are easy to overlook when formalizing complex specifications, demonstrating the value of executable testing for validating formal semantics.
We used the test results to correct both our correctness properties and instruction semantics.

\begin{table}[t]
\centering
\caption{RISC-V Test Suite Coverage}
\label{tab:test-coverage}
\begin{tabular}{lcccccr}
\toprule
 & \textbf{I} & \textbf{M} & \textbf{F} & \textbf{A} & \textbf{Zicsr} & \textbf{Total} \\
 & \multicolumn{6}{c}{\small(\# test files / \# test cases)} \\
\midrule
RV32 & 41 / 926   & 8 / 174   & 11 / 209  & 9 / 40   & 10 / 47 & 79 / 1,396 \\
RV64 & 52 / 1,354 & 13 / 221  & 10 / 177  & 18 / 80  & -- / -- & 93 / 1,832 \\
\midrule
\textbf{Total} & 93 / 2,280 & 21 / 395 & 21 / 386 & 27 / 120 & 10 / 47 & \textbf{172 / 3,228} \\
\bottomrule
\end{tabular}
\end{table}

\section{Case Studies}
\label{sec:evaluation}

We validate our \itree-based \riscv semantics through three case studies
spanning both sides of the ISA boundary:
cross-level bisimulation between LLVM IR and \riscv code via Vellvm (\S\ref{subsec:llvm-riscv}),
translation validation of a compiler reordering for macro-operation fusion (\S\ref{subsec:macro-fusion}),
and refinement of a \koika hardware ALU against our ISA specification (\S\ref{subsec:koika}).

\subsection{Cross-Level Bisimulation with Vellvm}
\label{subsec:llvm-riscv}

Since both our semantics and Vellvm~\cite{Zakowski2021LLVM} represent programs as ITrees, we can prove semantic equivalence between LLVM IR and \riscv code directly using ITree's heterogeneous bisimulation (\texttt{rutt}).
A verified compiler from LLVM IR to \riscv is outside the scope of this work; instead, we compare the \itree representations of an LLVM IR program and its corresponding \riscv code, establishing the proof infrastructure and demonstrating it on a concrete example.

\subsubsection{Proof Infrastructure}

\smallskip\noindent\textbf{Interpretation.}
LLVM IR and \riscv operate over different state representations---local variables versus registers---so their raw ITrees emit incompatible events.
The key idea is to apply interpreters on each side that handle these internal events, reducing both programs to ITrees that emit only observable memory events.
On the LLVM side, Vellvm's \texttt{interp\_cfg2} handles variable accesses internally, yielding an ITree with only observable \texttt{MemoryE} events (e.g., \texttt{Load}, \texttt{Store}).
On the \riscv side, our \texttt{interp\_state riscv\_handler} interprets \texttt{ProcessorE} events by reading/writing the register file and fetching instructions from memory, leaving only \texttt{VMemE} events (\texttt{VMemRead}, \texttt{VMemWrite}) observable.
After interpretation, both sides speak the same ``language'' of memory events, enabling direct comparison.

\smallskip\noindent\textbf{Four Relations.}
To prove bisimulation between any pair of LLVM IR and compiled \riscv code, one must instantiate four relations (Listing~\ref{lst:four-rel}):

\noindent\begin{minipage}{\linewidth}
\begin{lstlisting}[language=coq, caption={Four relations for cross-level bisimulation}, label={lst:four-rel}]
Definition obs_REv : <|$\forall$|> A B, MemoryE A <|$\to$|> VMemE B <|$\to$|> Prop :=
  fun A B e1 e2 =>
    match e1, e2 with
    | Load addr, VMemRead raddr offset width =>
        addr_eq addr (bv_add raddr offset) <|$\land$|> width = Word
    ... end.

Definition obs_RAns A B (e1 : MemoryE A) (a : A) (e2 : VMemE B) (b : B) : Prop :=
  is_ok b <|$\land$|> a = to_uvalue (unwrap b) <|$\land$|> ...

Definition init_rel (l : local_env) (st : state) : Prop :=
  lookup l "base" = to_uvalue (get_reg a0 st) <|$\land$|>
  lookup l "idx"  = to_uvalue (get_reg a1 st) <|$\land$|> ...

Definition final_rel (l' : local_env) (st' : state) : Prop :=
  lookup l' "v1" = to_uvalue (get_reg a2 st') <|$\land$|> ...
\end{lstlisting}
\end{minipage}

\begin{enumerate}[leftmargin=*,nosep]
\item \emph{Event relation} (\texttt{obs\_REv}): relates LLVM \texttt{MemoryE} events to \riscv \texttt{VMemE} events.
\texttt{addr\_eq} asserts that the LLVM \texttt{Load} address equals the \riscv address \texttt{bv\_add raddr offset}---LLVM uses a single computed address, whereas \riscv uses base-plus-offset addressing---and the \riscv access width must be \texttt{Word}, matching LLVM's default word-sized load.
\item \emph{Response relation} (\texttt{obs\_RAns}): requires the \riscv access to succeed (\texttt{is\_ok b}) and the response values to agree (\texttt{a = to\_uvalue (unwrap b)}). The success guard reflects that LLVM semantics lacks memory error modeling, so bisimulation holds only for valid accesses.
\item \emph{Initial/final relations} (\texttt{init\_rel}, \texttt{final\_rel}): relate LLVM local environments to \riscv register states, and memory states on both sides, as preconditions and postconditions.
For instance, \texttt{init\_rel} maps LLVM variables \texttt{"base"} and \texttt{"idx"} to registers \texttt{a0} and \texttt{a1}, while \texttt{final\_rel} maps the result variable \texttt{"v1"} to \texttt{a2}.
\end{enumerate}

\smallskip\noindent\textbf{Bisimulation via \texttt{rutt}.}
Given these four relations, the proof obligation is a single \texttt{rutt} judgement, where \texttt{llvm\_prog} and \texttt{riscv\_prog} are the ITree representations of the LLVM IR and compiled \riscv code, $g$ and $l$ are the LLVM global and local environments, and $st$ is the \riscv processor state:
\begin{multline*}
\texttt{rutt}~\texttt{obs\_REv}~\texttt{obs\_RAns}~\texttt{final\_rel}\\
(\texttt{interp\_cfg2}~\texttt{llvm\_prog}~g~l)~(\texttt{interp\_state}~\texttt{riscv\_handler}~\texttt{riscv\_prog}~st)
\end{multline*}
The heterogeneous weak bisimulation \texttt{rutt} allows both sides to take differing numbers of internal $\tau$ steps while requiring that every observable event on one side is matched by a related event on the other.

\smallskip\noindent\textbf{Per-Instruction Step Lemmas.}
In the bisimulation proof, stepping through a \riscv instruction does not simply invoke the execution functions in Listing~\ref{lst:rtype}, such as \texttt{exec\_RTYPE}; we implement a \texttt{step} function that wraps the fetch, decode, and execution of each instruction, including PC reads, memory fetches, and decoding.
Directly symbolically executing through this cycle for every instruction in a bisimulation proof would be prohibitively slow.
Instead, for each instruction $i$ we prove a lemma that summarises the net effect of \texttt{step}~$i$ after interpretation: one fetch-decode-execute step updates the PC, integer registers, and floating-point registers, and, for memory instructions, emits the corresponding \texttt{VMemE} event.
These lemmas allow each instruction to be ``stepped through'' in a single rewrite, making bisimulation proofs tractable.

\smallskip\noindent\textbf{Generality.}
The infrastructure above is not specific to any particular code pattern.
For any LLVM IR program and its compiled \riscv counterpart, one instantiates the four relations according to the relevant memory operations and the register-variable mapping dictated by the calling convention.
The proof then proceeds uniformly: symbolically execute both interpreted ITrees and show they satisfy \texttt{rutt}.
This makes the approach a reusable framework for verifying individual compilation results, rather than a one-off proof.

\subsubsection{Example: Array Element Load}
We demonstrate the framework on a common memory access pattern---loading an element from an array given a base pointer and index.
In C, this corresponds to \texttt{v1 = base[idx]}.
Figure~\ref{fig:llvm-riscv-comparison} shows the correspondence between LLVM IR and compiled RISC-V code.
The LLVM IR computes the element address using \texttt{getelementptr} and loads the value.
The RISC-V code uses register \texttt{a0} for \texttt{base} and \texttt{a1} for \texttt{idx}:
\texttt{slli} computes the byte offset by shifting left by 2 (multiplying by 4 for 32-bit elements), \texttt{add} computes the element address, and \texttt{lw} loads the word from memory.

\begin{figure}[H]
\centering
\begin{minipage}{0.55\textwidth}
\centering
\textbf{LLVM IR} (\texttt{llvm\_array\_load})
\begin{lstlisting}[numbers=none]
%p1 = getelementptr i32, i32* %base, i32 %idx

%v1 = load i32, i32* %p1

\end{lstlisting}
\end{minipage}
\hfill
\begin{minipage}{0.35\textwidth}
\centering
\textbf{RISC-V} (\texttt{riscv\_array\_load})
\begin{lstlisting}[language=RISCV, numbers=none]
  slli t0, a1, 2
  add  t0, a0, t0
  lw   a2, 0(t0)
\end{lstlisting}
\end{minipage}
\caption{LLVM IR array load and compiled RISC-V code.}
\label{fig:llvm-riscv-comparison}
\end{figure}

\begin{figure}[H]
\centering
\hspace{-1.5cm}%
\begin{tikzpicture}[
    node distance=0.6cm,
    taunode/.style={circle, draw, fill=gray!20, minimum size=0.55cm, font=\small},
    visnode/.style={rectangle, draw, fill=blue!15, rounded corners, minimum width=1.6cm, minimum height=0.55cm, font=\small},
    retnode/.style={rectangle, draw, fill=green!15, rounded corners, minimum width=1cm, minimum height=0.55cm, font=\small},
    every edge/.style={draw, ->, >=stealth}
]

\node[font=\bfseries] (llvm-title) at (-1.2, 0) {LLVM};
\node[taunode, right=0.4cm of llvm-title] (l-tau1) {$\tau$};
\node[visnode, right=0.8cm of l-tau1] (l-load) {Load addr (\texttt{load})};
\node[taunode, right=0.8cm of l-load] (l-tau2) {$\tau$};
\node[retnode, right=0.8cm of l-tau2] (l-ret) {Ret};

\draw[->] (l-tau1) -- (l-load) node[midway, above, font=\scriptsize] {gep};
\draw[->] (l-load) -- (l-tau2) node[midway, above, font=\scriptsize] {val};
\draw[->] (l-tau2) -- (l-ret);

\node[font=\bfseries] (rv-title) at (-3.5, -1.6) {RISC-V};
\node[visnode] (r-load) at (l-load |- rv-title) {VMemRead addr (\texttt{lw})};
\node[taunode, left=0.8cm of r-load] (r-tau2) {$\tau$};
\node[taunode, left=1.0cm of r-tau2] (r-tau1) {$\tau$};
\node[taunode, right=0.7cm of r-load] (r-tau3) {$\tau$};
\node[retnode, right=0.7cm of r-tau3] (r-ret) {Ret};

\draw[->] (r-tau1) -- (r-tau2) node[midway, above, font=\scriptsize] {slli};
\draw[->] (r-tau2) -- (r-load) node[midway, above, font=\scriptsize] {add};
\draw[->] (r-load) -- (r-tau3) node[midway, above, font=\scriptsize] {val};
\draw[->] (r-tau3) -- (r-ret);

\draw[dashed, red, thick, <->] (l-load.south) -- (r-load.north) node[midway, right, font=\small\bfseries] {rutt};

\end{tikzpicture}
\caption{ITree structures after interpretation.}
\label{fig:itree-bisim}
\end{figure}
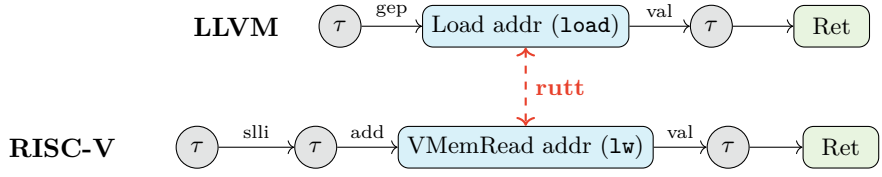

In Listing~\ref{lst:four-rel}, \texttt{init\_rel} maps LLVM variables \texttt{base} and \texttt{idx} to registers \texttt{a0} and \texttt{a1} following the calling convention, and \texttt{final\_rel} requires the LLVM result \texttt{v1} to match register \texttt{a2}.
The bisimulation theorem (Listing~\ref{lst:bisim-thm}) states that given compatible initial states satisfying \texttt{init\_rel} and a \texttt{no\_overflow} assumption (ensuring address computations stay within the 32-bit address space), the two interpreted programs are bisimilar under \texttt{rutt}.

\noindent\begin{minipage}{\linewidth}
\begin{lstlisting}[language=coq, caption={LLVM-RISC-V weak bisimulation theorem}, label={lst:bisim-thm}]
Theorem llvm_riscv_weak_bisim :
  <|$\forall$|> (g : global_env) (l : local_env) (st : state),
  init_rel l st <|$\to$|> no_overflow st <|$\to$|>
  rutt obs_REv obs_RAns final_rel (interp_cfg2 llvm_array_load g l)
                                   (interp_state riscv_handler riscv_array_load st).
\end{lstlisting}
\end{minipage}

Figure~\ref{fig:itree-bisim} illustrates the proof structure.
After interpretation, the LLVM side reduces to internal $\tau$ steps for address computation (\texttt{getelementptr}), followed by a visible \texttt{Load} event.
The \riscv side reduces to $\tau$ steps for \texttt{slli} and \texttt{add} (register operations resolved by the handler), followed by a visible \texttt{VMemRead} event.
The proof applies the per-instruction step lemmas to step through each \riscv instruction in a single rewrite, then uses \texttt{rutt} coinduction to show that the two visible events satisfy \texttt{obs\_REv}, their responses satisfy \texttt{obs\_RAns}, and the final states satisfy \texttt{final\_rel}.

\subsection{Translation Validation for Macro-Operation Fusion Optimization}
\label{subsec:macro-fusion}

Macro-Operation Fusion allows hardware to combine adjacent
instruction pairs (e.g., \inst{lui}/\inst{addi},
\inst{auipc}/\inst{jalr}) into single micro-operations ($\mu$ops), reducing
frontend pressure.
However, standard compiler schedulers actively separate dependent
pairs to hide latency, breaking fusion opportunities.
Restoring adjacency requires instruction reordering, which must
preserve data dependencies and, for PC-relative instructions like
\inst{auipc}, account for changes in memory position.
We use translation validation to prove that such reordering
preserves semantic equivalence.

\begin{wrapfigure}{r}{0.52\linewidth}
\vspace{0em}
\begin{lstlisting}[language=RISCV, caption={Baseline: Fusion Inhibited}, label={lst:baseline}]
lui   a0, 0x80000    # a0 = 0x80000000
sub   t0, t1, t2     # t0 = t1 - t2
auipc ra, 0x0        # ra = PC + 0
addi  a0, a0, 0x456  # a0 = 0x80000456
jalr  ra, 100(ra)    # jump to ra+100
\end{lstlisting}
\vspace{-1.5em}
\end{wrapfigure}

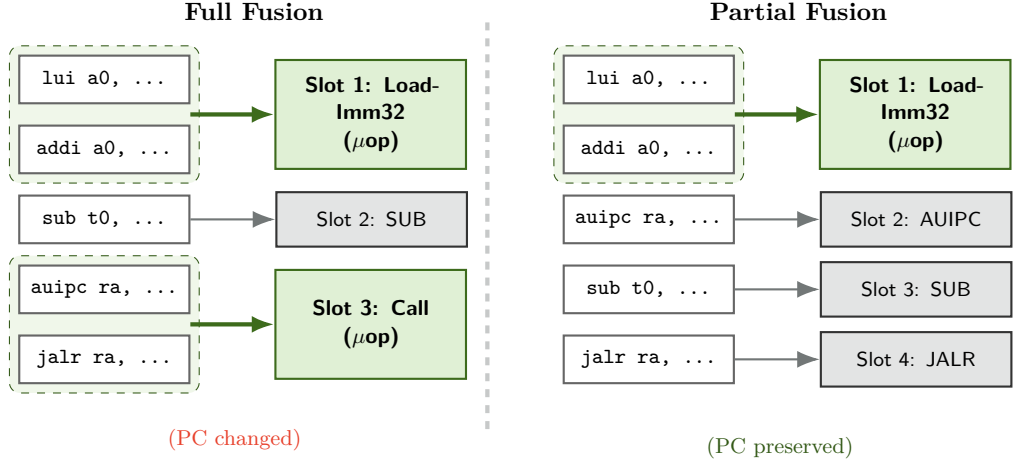
\begin{figure}[H]
    \centering
    \begin{tikzpicture}[node distance=0.3cm, scale=0.9, every node/.style={transform shape}]
        \begin{scope}[local bounding box=fullfusion]
            \node[font=\bfseries\large] (ttl1) at (2.2, 1) {Full Fusion};

            \node[inst] (i1) at (0,0) {lui a0, \dots};
            \node[inst, below=of i1] (i2) {addi a0, \dots};
            \node[inst, below=of i2] (i3) {sub t0, \dots};
            \node[inst, below=of i3] (i4) {auipc ra, \dots};
            \node[inst, below=of i4] (i5) {jalr ra, \dots};

            \begin{pgfonlayer}{background}
                \node[fit=(i1)(i2), fill=green!10, draw=green!50!black, dashed, rounded corners] {};
                \node[fit=(i4)(i5), fill=green!10, draw=green!50!black, dashed, rounded corners] {};
            \end{pgfonlayer}

            \node[fusedslot, right=2.5cm of $(i1)!0.5!(i2)$, text width=2.5cm] (fs1) {\textbf{Slot 1: LoadImm32}\\($\mu$op)};
            \node[slot, right=1.25cm of i3, text width=2.5cm] (fs2) {Slot 2: SUB};
            \node[fusedslot, right=2.5cm of $(i4)!0.5!(i5)$, text width=2.5cm] (fs3) {\textbf{Slot 3: Call}\\($\mu$op)};

            \draw[fusionarrow] ($(i1.east)!0.5!(i2.east)$) -- (fs1);
            \draw[arrow] (i3) -- (fs2);
            \draw[fusionarrow] ($(i4.east)!0.5!(i5.east)$) -- (fs3);

            \node[below=of fs3, font=\small, xshift=-2cm, yshift=-0.3cm, red] {(PC changed)};
        \end{scope}

        \draw[gray!40, dashed, ultra thick] ($(fullfusion.east)+(0.3cm, 3)$) -- ($(fullfusion.east)+(0.3cm, -3)$);

        \begin{scope}[local bounding box=partial, xshift=8cm]
             \node[font=\bfseries\large] (ttl2) at (2.2, 1) {Partial Fusion};

            \node[inst] (oi1) at (0,0) {lui a0, \dots};
            \node[inst, below=of oi1] (oi2) {addi a0, \dots};
            \node[inst, below=of oi2] (oi3) {auipc ra, \dots};
            \node[inst, below=of oi3] (oi4) {sub t0, \dots};
            \node[inst, below=of oi4] (oi5) {jalr ra, \dots};

            \begin{pgfonlayer}{background}
                \node[fit=(oi1)(oi2), fill=green!10, draw=green!50!black, dashed, rounded corners] {};
            \end{pgfonlayer}

            \node[fusedslot, right=2.5cm of $(oi1)!0.5!(oi2)$, text width=2.5cm] (pfs1) {\textbf{Slot 1: LoadImm32}\\($\mu$op)};
            \node[slot, right=1.25cm of oi3, text width=2.5cm] (pfs2) {Slot 2: AUIPC};
            \node[slot, right=1.25cm of oi4, text width=2.5cm] (pfs3) {Slot 3: SUB};
            \node[slot, right=1.25cm of oi5, text width=2.5cm] (pfs4) {Slot 4: JALR};

            \draw[fusionarrow] ($(oi1.east)!0.5!(oi2.east)$) -- (pfs1);
            \draw[arrow] (oi3) -- (pfs2);
            \draw[arrow] (oi4) -- (pfs3);
            \draw[arrow] (oi5) -- (pfs4);

             \node[below=of pfs4, font=\small, xshift=-2cm, yshift=-0.3cm, green!50!black] {(PC preserved)};
        \end{scope}

    \end{tikzpicture}
    \caption{Comparison of fusion strategies.}
    \label{fig:fusion_diagram}
\end{figure}

\smallskip\noindent\textbf{Baseline.}
A classic latency-prioritized compiler scheduler produces the
ordering in Listing~\ref{lst:baseline}: a function call sequence
with two fusible pairs---\inst{lui}/\inst{addi} (fusible to
\textbf{LoadImm32}) and \inst{auipc}/\inst{jalr} (fusible to
\textbf{CallImm}).
The scheduler interleaves an independent \inst{sub} between the
pairs to hide latency, but this breaks adjacency and prevents
fusion, requiring all \textbf{5 $\mu$ops} to be decoded
separately.
Figure~\ref{fig:fusion_diagram} compares two reordering strategies:

\textbf{Full Fusion (Semantics-Breaking):}
Placing both \inst{lui}/\inst{addi} and \inst{auipc}/\inst{jalr} pairs adjacently achieves maximum fusion, reducing 5 instructions to \textbf{3 $\mu$ops}. However, this reordering moves \inst{auipc} to a different memory address. Since \inst{auipc} computes $ra = \texttt{PC}$, changing its position changes the jump target---a correctness violation in Position-Independent Code.

\textbf{Partial Fusion (Semantics-Preserving):}
The correct optimization keeps \inst{auipc} at its original position while grouping only \inst{lui}/\inst{addi} adjacently. This enables \textbf{LoadImm32} fusion, yielding \textbf{4 $\mu$ops}. The \inst{sub} instruction moves after \inst{auipc} but before \inst{jalr}---a safe reordering since \inst{sub} has no data dependencies with the call sequence.

We apply translation validation to formally prove that the partial fusion reordering preserves program semantics.
The equivalence theorem (Listing~\ref{lst:fusion_equiv}) is parameterized over symbolic immediates (\texttt{imm\_hi}, \texttt{imm\_lo}, \texttt{pc\_off}, \texttt{call\_off}) rather than the concrete values in the case study (e.g., \texttt{0x80000}, \texttt{0x456}), ensuring the equivalence holds for any valid instruction encoding.
\texttt{baseline\_seq} encodes the original instruction sequence (Listing~\ref{lst:baseline}) as an ITree, and \texttt{partial\_fused\_seq} encodes the reordered sequence after the partial fusion pass (Figure~\ref{fig:fusion_diagram}, right).
The \texttt{state\_eq} precondition requires that both initial states have identical register values and memory contents.
The \texttt{result\_equiv} relation ensures that final states agree on all registers modified by the instruction sequence (\texttt{a0}, \texttt{t0}, \texttt{ra}, and \texttt{pc}).
Crucially, \texttt{pc} equivalence holds because \inst{auipc} remains at its original position in both sequences, computing the same PC-relative address.
In contrast, a \texttt{full\_fused\_seq} that moves \inst{auipc} would fail to satisfy \texttt{result\_equiv}: the relocated \inst{auipc} computes a different \texttt{ra} value, causing \inst{jalr} to jump to a different target address.
The proof proceeds by symbolically executing both sequences using the per-instruction step lemmas (\S\ref{subsec:llvm-riscv}), which reduce each \texttt{step} to its net state update in a single rewrite.
After stepping through all instructions on both sides, the proof discharges \texttt{result\_equiv} by showing that the final register values coincide.

\noindent\begin{minipage}{\linewidth}
\begin{lstlisting}[language=coq, caption={Semantic equivalence of instruction reordering}, label={lst:fusion_equiv}, morekeywords={Theorem}]
Theorem fusion_reorder_equiv :
  <|$\forall$|> imm_hi imm_lo pc_off call_off st_orig st_reord,
  let baseline := baseline_seq imm_hi imm_lo pc_off call_off in
  let fused := partial_fused_seq imm_hi imm_lo pc_off call_off in
  state_eq st_orig st_reord <|$\to$|>
  eutt result_equiv (interp_state riscv_handler baseline st_orig)
                    (interp_state riscv_handler fused st_reord).
\end{lstlisting}
\end{minipage}

\subsection{Refinement of \koika ALU Against ISA Specification}
\label{subsec:koika}

\textsc{\koika}~\cite{Bourgeat2020Koika} is a rule-based hardware description
language with mechanised semantics in \rocq, which also provides a
\riscv processor implementation.
We prove that a \textsc{\koika} ALU design computes the same
result as our ISA-level \itree model for all ten R-type integer
operations.
\textsc{\koika} provides an interpreter
(\texttt{interp\_action}) that evaluates typed circuit
descriptions within a register environment.
We wrap it into \texttt{run\_execALU32}, which takes a 32-bit
encoded instruction and operand values and returns the ALU result.

We prove the correctness lemma (Listing~\ref{lst:alu-refine}) that
connects the two levels through an existential witness~\texttt{v}.
The first conjunct states that running the R-type instruction
in our ITree ISA model produces a state where the destination
register \texttt{rd} holds value~\texttt{v}.
The second conjunct states that the \textsc{\koika} ALU,
given the same encoded operation and operand values read from
the state, produces the bit-level representation of the same
value~\texttt{v}---which will ultimately be committed to
register \texttt{rd}.
Thus \texttt{v} serves as the connecting witness between the
ISA specification and the hardware design, ensuring they
agree on the value written to the destination register.

\noindent\begin{minipage}{\linewidth}
\begin{lstlisting}[language=coq, caption={Refinement of \koika ALU against ISA specification}, label={lst:alu-refine}, morekeywords={Theorem}]
Theorem exec_RTYPE_execALU32_consistent :
  <|$\forall$|> op rs1 rs2 rd s, rd <|$\neq$|> x0 <|$\to$|> <|$\exists$|> v,
    interp_state handler                (* ISA (ITree) *)
      (exec_RTYPE rs1 rs2 rd op) s
      <|$\approx$|> Ret (set_reg s rd v, Success)
  <|$\wedge$|> run_execALU32 (encode op)          (* Hardware (Koika) *)
      (get_reg s rs1) (get_reg s rs2) = Some (to_bits v).
\end{lstlisting}
\end{minipage}

Collectively, the three case studies exercise the semantics across the abstraction stack: upward against a compiler IR (Vellvm), laterally across microarchitectural reorderings, and downward against a hardware design (\koika).
The same per-instruction step lemmas and event-based reasoning underpin all three proofs, confirming that a single ITree-based semantics can serve as a shared reference point across the hardware--software boundary.



\section{Related Work}
\label{sec:related}

Our work builds on three lines of research: ISA specification languages, Interaction Trees, and verified compilation and hardware refinement.

\emph{ISA Specification Languages.}
Several domain-specific languages have been developed for ISA specification.
The Sail language~\cite{Armstrong2019Sail} provides a first-order imperative language for defining ISA semantics, supporting automatic generation of executable emulators and theorem prover definitions for Isabelle, HOL4, and Rocq.
The Sail RISC-V model has been adopted as an official reference by RISC-V International and serves as the basis for the Isla symbolic execution engine~\cite{Armstrong2021Isla} and the Islaris machine code verification framework~\cite{Sammler2022Islaris}.
While Sail provides excellent coverage of ISA features, the generated Rocq code requires external trusted translation, and the sequential imperative style limits compositional reasoning.

Fox~\cite{Fox2012L3} proposes a monadic functional style for ISA specification, successfully applied to ARMv7~\cite{Fox2010ARMv7} and other architectures.
MiniSail~\cite{MiniSail2021AFP} provides a formalized kernel of the Sail language in Isabelle/HOL.
Kanabar et al.~\cite{Kanabar2022Taming} demonstrated validation of an authoritative ARMv8 ISA specification and its use for CakeML compiler verification.
The MIT PLV group developed RISC-V specifications in Haskell using abstract monads~\cite{Bourgeat2023RiscvMonads}, translated to Rocq via hs-to-coq. However, the typeclass-parameterized monadic specification does not expose coinductive structure, which precludes bisimulation-based reasoning across abstraction levels. Our approach defines semantics directly in Rocq using ITrees, providing native support for weak bisimulation and refinement over potentially nonterminating computations.

\emph{Interaction Tree Applications.}
Since their introduction~\cite{Xia2020ITrees}, Interaction Trees have been applied to diverse verification tasks.
The Vellvm project~\cite{Zakowski2021LLVM,Zhao2012Vellvm} provides a modular, compositional semantics for LLVM IR, enabling verification of compiler optimizations and serving as a foundation for our LLVM-to-RISC-V equivalence proofs.
ITrees have been used to specify and verify networked servers in C~\cite{Koh2019DeepWeb,Li2021CertiCoq}, demonstrating their applicability to systems with complex I/O behavior.
The HELIX project~\cite{Zaliva2020HelixTranslation} uses the
\itree-based Vellvm backend to verify the compilation of
high-performance numerical DSLs down to LLVM IR in the SPIRAL
framework.

Several extensions address limitations of the original framework.
Choice Trees~\cite{Chappe2023ChoiceTrees} extend ITrees with nondeterministic choice, applied to concurrent memory models~\cite{Chappe2025MonadicInterpreters}.
Guarded Interaction Trees~\cite{Frumin2024GITrees} support higher-order effects within the Iris framework, with extensions for context-dependent effects~\cite{Stepanenko2025ContextGITrees}.
HITrees~\cite{Ayyam2025HITrees} provide an alternative approach to higher-order effects using defunctionalization.
The ITree framework has been ported to Isabelle/HOL~\cite{Foster2024IsabelleITrees} and extended for modular program logics~\cite{Vistrup2025ProgramLogics} and security properties~\cite{Silver2023Noninterference}.

\emph{Verified Compilation and Hardware Refinement.}
Verified compilation has been pioneered by CompCert~\cite{Leroy2009CACM,Leroy2006POPL,Leroy2009JAR}, which provides a fully verified C compiler targeting multiple architectures, including \riscv.
However, its ISA semantics are internal to the compiler and not exposed as a reusable specification for hardware refinement or cross-level verification.
CakeML~\cite{Kumar2014CakeML,Tan2016CakeMLBackend,Tan2019JFP} provides a verified ML compiler with proofs extending to machine code.
Translation validation~\cite{Pnueli1998TranslationValidation,Necula2000TranslationValidation} offers an alternative approach that validates compiler outputs rather than verifying the compiler itself.
Beyond semantic preservation, complementary work has explored preserving formal properties proven at the source level across compilation~\cite{wppresPriSC}.

For hardware verification, Kami~\cite{Choi2017Kami} provides a platform for parametric hardware specification and modular verification in Rocq.
\koika~\cite{Bourgeat2020Koika,PitClaudel2021Cuttlesim} captures the essence of Bluespec for rule-based hardware design with efficient simulation.
Bourgeat et al.~\cite{Bourgeat2025Fjfj} address concurrent hardware verification.
The integration of software and hardware verification has been demonstrated for a simple embedded system~\cite{Erbsen2021Integration}, connecting the Bedrock2 compiler and the Kami processor through a shared \riscv specification parameterized via Haskell type classes~\cite{Bourgeat2023RiscvMonads}.
None of the above hardware verification frameworks use \itree{}-based property specifications. In contrast, \itrees natively support bisimulation and refinement, and can naturally express advanced security properties such as noninterference~\cite{Silver2023Noninterference}, opening the door to verifying hardware against non-functional specifications beyond correctness.

\section{Conclusion}
\label{sec:conclusion}

We presented an \itree{}-based formal semantics of \riscv in
\rocq that serves as a shared foundation for verification on both
sides of the ISA boundary.
Three case studies validate this: cross-level bisimulation between
LLVM IR and \riscv code, translation validation of macro-operation
fusion, and refinement of a \koika hardware ALU against the ISA
specification.

Future work aims to scale toward end-to-end correctness:
proving LLVM lowering correct for general programs, verifying a
real-world \riscv processor implementation against our ISA
specification, and extracting SystemVerilog assertions from our
formalization for use in hardware model checking.

\bibliography{ref,../references}

\begin{thebibliography}{10}

\bibitem{wppresPriSC}
Carmine Abate, Mohamed Elsheikh, Kleio Liotati, Franti\v{s}ek Farka, and
  Sebastian Ertel.
\newblock {WP}-preserving compilation---preserving weakest preconditions for
  end-to-end verification.
\newblock In {\em Principles of Secure Compilation (PriSC 2026)}, 2026.

\bibitem{Armstrong2019Sail}
Alasdair Armstrong, Thomas Bauereiss, Brian Campbell, Alastair Reid, Kathryn~E.
  Gray, Robert~M. Norton, Prashanth Mundkur, Mark Wassell, Jon French,
  Christopher Pulte, Shaked Flur, Ian Stark, Neel Krishnaswami, and Peter
  Sewell.
\newblock {ISA} semantics for {ARMv8-A}, {RISC-V}, and {CHERI-MIPS}.
\newblock {\em Proc. ACM Program. Lang.}, 3(POPL), 2019.
\newblock \href {https://doi.org/10.1145/3290384} {\path{doi:10.1145/3290384}}.

\bibitem{Armstrong2021Isla}
Alasdair Armstrong, Brian Campbell, Ben Simner, Christopher Pulte, and Peter
  Sewell.
\newblock Isla: Integrating full-scale {ISA} semantics and axiomatic
  concurrency models.
\newblock In {\em Proceedings of the 33rd International Conference on
  Computer-Aided Verification (CAV 2021)}, volume 12759 of {\em Lecture Notes
  in Computer Science}, pages 303--316. Springer, 2021.
\newblock \href {https://doi.org/10.1007/978-3-030-81685-8_14}
  {\path{doi:10.1007/978-3-030-81685-8_14}}.

\bibitem{ArranzOlmos2025Jasmin}
Santiago Arranz-Olmos, Gilles Barthe, Lionel Blatter, Benjamin Gr{\'e}goire,
  Vincent Laporte, and Paolo Torrini.
\newblock The jasmin compiler preserves cryptographic security.
\newblock 2025.
\newblock \href {https://arxiv.org/abs/2511.11292} {\path{arXiv:2511.11292}},
  \href {https://doi.org/10.48550/arXiv.2511.11292}
  {\path{doi:10.48550/arXiv.2511.11292}}.

\bibitem{Asanovic2016RocketChip}
Krste Asanovi{\'c}, Rimas Avizienis, Jonathan Bachrach, Scott Beamer, David
  Biancolin, Christopher Celio, Henry Cook, Daniel Dabbelt, John Hauser, Adam
  Izraelevitz, Sagar Karandikar, Ben Keller, Donggyu Kim, John Koenig, Yunsup
  Lee, Eric Love, Martin Maas, Albert Magyar, Howard Mao, Miquel Moreto, Albert
  Ou, David~A. Patterson, Brian Richards, Colin Schmidt, Stephen Twigg, Huy Vo,
  and Andrew Waterman.
\newblock The rocket chip generator.
\newblock Technical Report UCB/EECS-2016-17, EECS Department, University of
  California, Berkeley, April 2016.

\bibitem{Asanovic2014Free}
Krste Asanovi{\'c} and David~A. Patterson.
\newblock Instruction sets should be free: The case for {RISC-V}.
\newblock Technical Report UCB/EECS-2014-146, EECS Department, University of
  California, Berkeley, August 2014.

\bibitem{Ayyam2025HITrees}
Amir Mohammad~Fadaei Ayyam and Michael Sammler.
\newblock {HITrees}: Higher-order interaction trees.
\newblock {\em arXiv preprint arXiv:2510.14558}, 2025.
\newblock URL: \url{https://arxiv.org/abs/2510.14558}.

\bibitem{Bourgeat2023RiscvMonads}
Thomas Bourgeat, Ian Clester, Andres Erbsen, Samuel Gruetter, Pratap Singh,
  Andrew Wright, and Adam Chlipala.
\newblock Flexible instruction-set semantics via abstract monads (experience
  report).
\newblock {\em Proc. ACM Program. Lang.}, 7(ICFP), 2023.
\newblock \href {https://doi.org/10.1145/3607833} {\path{doi:10.1145/3607833}}.

\bibitem{bourgeat2021multipurpose}
Thomas Bourgeat, Ian Clester, Andres Erbsen, Samuel Gruetter, Andrew Wright,
  and Adam Chlipala.
\newblock A multipurpose formal risc-v specification.
\newblock {\em arXiv preprint arXiv:2104.00762}, 2021.

\bibitem{Bourgeat2025Fjfj}
Thomas Bourgeat, Jiazheng Liu, Adam Chlipala, and Arvind.
\newblock Making concurrent hardware verification sequential.
\newblock {\em Proc. ACM Program. Lang.}, 9(PLDI), 2025.
\newblock \href {https://doi.org/10.1145/3729331} {\path{doi:10.1145/3729331}}.

\bibitem{Bourgeat2020Koika}
Thomas Bourgeat, Cl{\'e}ment Pit-Claudel, Adam Chlipala, and Arvind.
\newblock The essence of bluespec: A core language for rule-based hardware
  design.
\newblock In {\em Proceedings of the 41st ACM SIGPLAN Conference on Programming
  Language Design and Implementation (PLDI 2020)}, pages 243--257. ACM, 2020.
\newblock \href {https://doi.org/10.1145/3385412.3385965}
  {\path{doi:10.1145/3385412.3385965}}.

\bibitem{Celio2015BOOM}
Christopher Celio, David~A. Patterson, and Krste Asanovi{\'c}.
\newblock The berkeley out-of-order machine ({BOOM}): An industry-competitive,
  synthesizable, parameterized {RISC-V} processor.
\newblock Technical Report UCB/EECS-2015-167, EECS Department, University of
  California, Berkeley, June 2015.

\bibitem{Chappe2023ChoiceTrees}
Nicolas Chappe, Paul He, Ludovic Henrio, Yannick Zakowski, and Steve Zdancewic.
\newblock Choice trees: Representing nondeterministic, recursive, and impure
  programs in {Coq}.
\newblock {\em Proc. ACM Program. Lang.}, 7(POPL), 2023.
\newblock \href {https://doi.org/10.1145/3571254} {\path{doi:10.1145/3571254}}.

\bibitem{Chappe2025MonadicInterpreters}
Nicolas Chappe, Ludovic Henrio, and Yannick Zakowski.
\newblock Monadic interpreters for concurrent memory models: Executable
  semantics of a concurrent subset of {LLVM IR}.
\newblock In {\em Proceedings of the 14th ACM SIGPLAN International Conference
  on Certified Programs and Proofs (CPP '25)}, pages 283--298. ACM, 2025.
\newblock \href {https://doi.org/10.1145/3703595.3705890}
  {\path{doi:10.1145/3703595.3705890}}.

\bibitem{Choi2017Kami}
Joonwon Choi, Muralidaran Vijayaraghavan, Benjamin Sherman, Adam Chlipala, and
  Arvind.
\newblock Kami: A platform for high-level parametric hardware specification and
  its modular verification.
\newblock {\em Proc. ACM Program. Lang.}, 1(ICFP), 2017.
\newblock \href {https://doi.org/10.1145/3110268} {\path{doi:10.1145/3110268}}.

\bibitem{Erbsen2021Integration}
Andres Erbsen, Samuel Gruetter, Joonwon Choi, Clark Wood, and Adam Chlipala.
\newblock Integration verification across software and hardware for a simple
  embedded system.
\newblock In {\em Proceedings of the 42nd ACM SIGPLAN International Conference
  on Programming Language Design and Implementation (PLDI 2021)}, pages
  604--619. ACM, 2021.
\newblock \href {https://doi.org/10.1145/3453483.3454065}
  {\path{doi:10.1145/3453483.3454065}}.

\bibitem{Foster2024IsabelleITrees}
Simon Foster, Chung-Kil Hur, and Jim Woodcock.
\newblock Unifying model execution and deductive verification with interaction
  trees in {Isabelle/HOL}.
\newblock {\em ACM Trans. Softw. Eng. Methodol.}, 34(4), 2024.
\newblock \href {https://doi.org/10.1145/3702981} {\path{doi:10.1145/3702981}}.

\bibitem{Fox2012L3}
Anthony C.~J. Fox.
\newblock Directions in {ISA} specification.
\newblock In {\em Interactive Theorem Proving (ITP 2012)}, volume 7406 of {\em
  Lecture Notes in Computer Science}, pages 338--344. Springer, 2012.
\newblock \href {https://doi.org/10.1007/978-3-642-32347-8_23}
  {\path{doi:10.1007/978-3-642-32347-8_23}}.

\bibitem{Fox2010ARMv7}
Anthony C.~J. Fox and Magnus~O. Myreen.
\newblock A trustworthy monadic formalization of the {ARMv7} instruction set
  architecture.
\newblock In {\em Interactive Theorem Proving (ITP 2010)}, volume 6172 of {\em
  Lecture Notes in Computer Science}, pages 243--258. Springer, 2010.
\newblock \href {https://doi.org/10.1007/978-3-642-14052-5_18}
  {\path{doi:10.1007/978-3-642-14052-5_18}}.

\bibitem{Frumin2024GITrees}
Dan Frumin, Amin Timany, and Lars Birkedal.
\newblock Modular denotational semantics for effects with guarded interaction
  trees.
\newblock {\em Proc. ACM Program. Lang.}, 8(POPL), 2024.
\newblock \href {https://doi.org/10.1145/3632854} {\path{doi:10.1145/3632854}}.

\bibitem{He2020GPaco}
Paul He, Li-yao Xia, Yannick Zakowski, and Steve Zdancewic.
\newblock An equational theory for weak bisimulation via generalized
  parameterized coinduction.
\newblock In {\em Proceedings of the 9th ACM SIGPLAN International Conference
  on Certified Programs and Proofs (CPP '20)}, pages 149--162. ACM, 2020.
\newblock \href {https://doi.org/10.1145/3372885.3373813}
  {\path{doi:10.1145/3372885.3373813}}.

\bibitem{Jung2018Iris}
Ralf Jung, Robbert Krebbers, Jacques-Henri Jourdan, Ale{\v{s}} Bizjak, Lars
  Birkedal, and Derek Dreyer.
\newblock Iris from the ground up: A modular foundation for higher-order
  concurrent separation logic.
\newblock {\em J. Funct. Program.}, 28:e20, 2018.
\newblock \href {https://doi.org/10.1017/S0956796818000151}
  {\path{doi:10.1017/S0956796818000151}}.

\bibitem{Kanabar2022Taming}
Hrutvik Kanabar, Anthony C.~J. Fox, and Magnus~O. Myreen.
\newblock Taming an authoritative {Armv8} {ISA} specification: {L3} validation
  and {CakeML} compiler verification.
\newblock In {\em 13th International Conference on Interactive Theorem Proving
  (ITP 2022)}, volume 237 of {\em LIPIcs}, pages 20:1--20:22. Schloss Dagstuhl
  -- Leibniz-Zentrum f{\"u}r Informatik, 2022.
\newblock \href {https://doi.org/10.4230/LIPIcs.ITP.2022.20}
  {\path{doi:10.4230/LIPIcs.ITP.2022.20}}.

\bibitem{Kanabar2023PureCake}
Hrutvik Kanabar, Samuel Vivien, Oskar Abrahamsson, Magnus~O. Myreen, Michael
  Norrish, Johannes~{\AA}man Pohjola, and Riccardo Zanetti.
\newblock {PureCake}: A verified compiler for a lazy functional language.
\newblock {\em Proc. ACM Program. Lang.}, 7(PLDI), 2023.
\newblock \href {https://doi.org/10.1145/3591259} {\path{doi:10.1145/3591259}}.

\bibitem{Koh2019DeepWeb}
Nicolas Koh, Yao Li, Yishuai Li, Li-yao Xia, Lennart Beringer, Wolf Honor{\'e},
  William Mansky, Benjamin~C. Pierce, and Steve Zdancewic.
\newblock From {C} to interaction trees: Specifying, verifying, and testing a
  networked server.
\newblock In {\em Proceedings of the 8th ACM SIGPLAN International Conference
  on Certified Programs and Proofs (CPP '19)}, pages 234--248. ACM, 2019.
\newblock \href {https://doi.org/10.1145/3293880.3294106}
  {\path{doi:10.1145/3293880.3294106}}.

\bibitem{Kumar2014CakeML}
Ramana Kumar, Magnus~O. Myreen, Michael Norrish, and Scott Owens.
\newblock {CakeML}: A verified implementation of {ML}.
\newblock In {\em Proceedings of the 41st Annual ACM SIGPLAN-SIGACT Symposium
  on Principles of Programming Languages (POPL 2014)}, pages 179--192. ACM,
  2014.
\newblock \href {https://doi.org/10.1145/2535838.2535841}
  {\path{doi:10.1145/2535838.2535841}}.

\bibitem{Leroy2006POPL}
Xavier Leroy.
\newblock Formal certification of a compiler back-end, or: Programming a
  compiler with a proof assistant.
\newblock In {\em Conference Record of the 33rd ACM SIGPLAN-SIGACT Symposium on
  Principles of Programming Languages (POPL 2006)}, pages 42--54. ACM, 2006.
\newblock \href {https://doi.org/10.1145/1111037.1111042}
  {\path{doi:10.1145/1111037.1111042}}.

\bibitem{Leroy2009CACM}
Xavier Leroy.
\newblock Formal verification of a realistic compiler.
\newblock {\em Communications of the ACM}, 52(7):107--115, 2009.
\newblock \href {https://doi.org/10.1145/1538788.1538814}
  {\path{doi:10.1145/1538788.1538814}}.

\bibitem{Leroy2009JAR}
Xavier Leroy.
\newblock A formally verified compiler back-end.
\newblock {\em Journal of Automated Reasoning}, 43(4):363--446, 2009.
\newblock \href {https://doi.org/10.1007/s10817-009-9155-4}
  {\path{doi:10.1007/s10817-009-9155-4}}.

\bibitem{Li2021CertiCoq}
Yishuai Li, Li-yao Xia, and Steve Zdancewic.
\newblock From {C} to interaction trees: Specifying, verifying, and testing a
  networked server (extended abstract).
\newblock In {\em Proceedings of the 10th ACM SIGPLAN International Conference
  on Certified Programs and Proofs (CPP '21)}. ACM, 2021.

\bibitem{Milner1989CCS}
Robin Milner.
\newblock {\em Communication and Concurrency}.
\newblock Prentice-Hall, 1989.

\bibitem{Necula2000TranslationValidation}
George~C. Necula.
\newblock Translation validation for an optimizing compiler.
\newblock In {\em Proceedings of the ACM SIGPLAN 2000 Conference on Programming
  Language Design and Implementation (PLDI 2000)}, pages 83--94. ACM, 2000.
\newblock \href {https://doi.org/10.1145/349299.349314}
  {\path{doi:10.1145/349299.349314}}.

\bibitem{Park1981Bisimulation}
David Park.
\newblock Concurrency and automata on infinite sequences.
\newblock In {\em Theoretical Computer Science, 5th GI-Conference}, volume 104
  of {\em Lecture Notes in Computer Science}, pages 167--183. Springer, 1981.
\newblock \href {https://doi.org/10.1007/BFb0017309}
  {\path{doi:10.1007/BFb0017309}}.

\bibitem{PitClaudel2021Cuttlesim}
Cl{\'e}ment Pit-Claudel, Thomas Bourgeat, Stella Lau, Arvind, and Adam
  Chlipala.
\newblock Effective simulation and debugging for a high-level hardware language
  using software compilers.
\newblock In {\em Proceedings of the 26th International Conference on
  Architectural Support for Programming Languages and Operating Systems (ASPLOS
  2021)}, pages 789--803. ACM, 2021.
\newblock \href {https://doi.org/10.1145/3445814.3446720}
  {\path{doi:10.1145/3445814.3446720}}.

\bibitem{Pnueli1998TranslationValidation}
Amir Pnueli, Michael Siegel, and Eli Singerman.
\newblock Translation validation.
\newblock In {\em Tools and Algorithms for the Construction and Analysis of
  Systems (TACAS '98)}, volume 1384 of {\em Lecture Notes in Computer Science},
  pages 151--166. Springer, 1998.
\newblock \href {https://doi.org/10.1007/BFb0054170}
  {\path{doi:10.1007/BFb0054170}}.

\bibitem{RISCVTests}
{RISC-V Software}.
\newblock {RISC-V} tests: Unit tests for {RISC-V} processors.
\newblock \url{https://github.com/riscv-software-src/riscv-tests}, 2024.
\newblock Accessed: 2024.

\bibitem{Sammler2022Islaris}
Michael Sammler, Angus Hammond, Rodolphe Lepigre, Brian Campbell, Jean
  Pichon-Pharabod, Derek Dreyer, Deepak Garg, and Peter Sewell.
\newblock Islaris: Verification of machine code against authoritative {ISA}
  semantics.
\newblock In {\em Proceedings of the 43rd ACM SIGPLAN International Conference
  on Programming Language Design and Implementation (PLDI 2022)}, pages
  825--840. ACM, 2022.
\newblock \href {https://doi.org/10.1145/3519939.3523434}
  {\path{doi:10.1145/3519939.3523434}}.

\bibitem{Sangiorgi2009Origins}
Davide Sangiorgi.
\newblock On the origins of bisimulation and coinduction.
\newblock {\em ACM Transactions on Programming Languages and Systems}, 31(4),
  2009.
\newblock \href {https://doi.org/10.1145/1516507.1516510}
  {\path{doi:10.1145/1516507.1516510}}.

\bibitem{Shen2024MacroFusion}
Jian-Yu Shen and Shih-Wei Liao.
\newblock Evaluating and enhancing performance through macro-op fusion
  optimization with risc-v.
\newblock In {\em Workshop Proceedings of the 53rd International Conference on
  Parallel Processing}, ICPP Workshops '24, page 33–37, New York, NY, USA,
  2024. Association for Computing Machinery.
\newblock \href {https://doi.org/10.1145/3677333.3678150}
  {\path{doi:10.1145/3677333.3678150}}.

\bibitem{Silver2023Noninterference}
Lucas Silver, Paul He, Ethan Cecchetti, Andrew~K. Hirsch, and Steve Zdancewic.
\newblock Semantics for noninterference with interaction trees.
\newblock In {\em 37th European Conference on Object-Oriented Programming
  (ECOOP 2023)}, volume 263 of {\em LIPIcs}, pages 29:1--29:29. Schloss
  Dagstuhl -- Leibniz-Zentrum f{\"u}r Informatik, 2023.
\newblock \href {https://doi.org/10.4230/LIPIcs.ECOOP.2023.29}
  {\path{doi:10.4230/LIPIcs.ECOOP.2023.29}}.

\bibitem{Stepanenko2025ContextGITrees}
Sergei Stepanenko, Emma Nardino, Virgil Marionneau, Dan Frumin, Amin Timany,
  and Lars Birkedal.
\newblock Context-dependent effects and concurrency in guarded interaction
  trees.
\newblock {\em arXiv preprint arXiv:2512.11577}, 2025.
\newblock URL: \url{https://arxiv.org/abs/2512.11577}.

\bibitem{Tan2016CakeMLBackend}
Yong~Kiam Tan, Magnus~O. Myreen, Ramana Kumar, Anthony Fox, Scott Owens, and
  Michael Norrish.
\newblock A new verified compiler backend for {CakeML}.
\newblock In {\em Proceedings of the 21st ACM SIGPLAN International Conference
  on Functional Programming (ICFP 2016)}, pages 60--73. ACM, 2016.
\newblock \href {https://doi.org/10.1145/2951913.2951924}
  {\path{doi:10.1145/2951913.2951924}}.

\bibitem{Tan2019JFP}
Yong~Kiam Tan, Magnus~O. Myreen, Ramana Kumar, Anthony Fox, Scott Owens, and
  Michael Norrish.
\newblock The verified {CakeML} compiler backend.
\newblock {\em Journal of Functional Programming}, 29:e2, 2019.
\newblock \href {https://doi.org/10.1017/S0956796818000229}
  {\path{doi:10.1017/S0956796818000229}}.

\bibitem{Vistrup2025ProgramLogics}
Max Vistrup, Michael Sammler, and Ralf Jung.
\newblock Program logics {\`a} la carte.
\newblock {\em Proc. ACM Program. Lang.}, 9(POPL), 2025.
\newblock \href {https://doi.org/10.1145/3704847} {\path{doi:10.1145/3704847}}.

\bibitem{MiniSail2021AFP}
Mark Wassell.
\newblock {MiniSail} - a kernel language for the {ISA} specification language
  {SAIL}.
\newblock {\em Archive of Formal Proofs}, June 2021.
\newblock \url{https://isa-afp.org/entries/MiniSail.html}.

\bibitem{RISCVUnpriv2019}
Andrew Waterman and Krste Asanovi{\'c}.
\newblock {\em The {RISC-V} Instruction Set Manual, Volume {I}: User-Level
  {ISA}, Version 2.2}.
\newblock RISC-V Foundation, 2019.

\bibitem{RISCVPriv2021}
Andrew Waterman, Krste Asanovi{\'c}, and John Hauser.
\newblock {\em The {RISC-V} Instruction Set Manual, Volume {II}: Privileged
  Architecture}.
\newblock RISC-V International, 2021.
\newblock Document Version 20211203.

\bibitem{Waterman2016Design}
Andrew~Shell Waterman.
\newblock {\em Design of the {RISC-V} Instruction Set Architecture}.
\newblock PhD thesis, University of California, Berkeley, January 2016.
\newblock Technical Report UCB/EECS-2016-1.

\bibitem{Xia2020ITrees}
Li-yao Xia, Yannick Zakowski, Paul He, Chung-Kil Hur, Gregory Malecha,
  Benjamin~C. Pierce, and Steve Zdancewic.
\newblock Interaction trees: Representing recursive and impure programs in
  {Coq}.
\newblock {\em Proc. ACM Program. Lang.}, 4(POPL), 2020.
\newblock \href {https://doi.org/10.1145/3371119} {\path{doi:10.1145/3371119}}.

\bibitem{Zakowski2021LLVM}
Yannick Zakowski, Calvin Beck, Irene Yoon, Ilia Zaichuk, Vadim Zaliva, and
  Steve Zdancewic.
\newblock Modular, compositional, and executable formal semantics for {LLVM
  IR}.
\newblock {\em Proc. ACM Program. Lang.}, 5(ICFP), 2021.
\newblock \href {https://doi.org/10.1145/3473572} {\path{doi:10.1145/3473572}}.

\bibitem{Zaliva2020HelixTranslation}
Vadim Zaliva, Ilia Zaichuk, and Franz Franchetti.
\newblock Verified translation between purely functional and imperative domain
  specific languages in {HELIX}.
\newblock In {\em Formal Methods - 23rd International Symposium (FM 2020)},
  volume 12546 of {\em Lecture Notes in Computer Science}, pages 37--54.
  Springer, 2020.
\newblock \href {https://doi.org/10.1007/978-3-030-63618-0_3}
  {\path{doi:10.1007/978-3-030-63618-0_3}}.

\bibitem{Zhao2012Vellvm}
Jianzhou Zhao, Santosh Nagarakatte, Milo M.~K. Martin, and Steve Zdancewic.
\newblock Formalizing the {LLVM} intermediate representation for verified
  program transformations.
\newblock In {\em Proceedings of the 39th Annual ACM SIGPLAN-SIGACT Symposium
  on Principles of Programming Languages (POPL 2012)}, pages 427--440. ACM,
  2012.
\newblock \href {https://doi.org/10.1145/2103656.2103709}
  {\path{doi:10.1145/2103656.2103709}}.

\end{thebibliography}

\end{document}